\documentclass[onecolumn,amsmath,amssymb]{revtex4-1}
\usepackage{xcolor}
\usepackage{booktabs}
\usepackage{threeparttable}
\usepackage{dcolumn}
\usepackage{bm}
\usepackage{amsmath}
\usepackage{graphicx}
\usepackage{color}
\usepackage{lineno}
\usepackage[colorlinks,linkcolor=red,citecolor=blue,urlcolor=blue]{hyperref}
\usepackage{etoolbox}

\begin{document}
\title{Anomalous Localization and Mobility Edges in Non-Hermitian Quasicrystals with Disordered Imaginary Gauge Fields}
\author{Guolin Nan}
\affiliation{Sanli Honors College, School of Physics and Electronics Engineering, Shanxi University, Taiyuan 030006, China}
\author{Zhijian Li}
\affiliation{Sanli Honors College, School of Physics and Electronics Engineering, Shanxi University, Taiyuan 030006, China}
\affiliation{Institute of Theoretical Physics and State Key Laboratory of Quantum Optics Technologies and Devices, Shanxi University, Taiyuan 030006, China}
\author{Feng Mei}
\affiliation{State Key Laboratory of Quantum Optics Technologies and Devices, Institute of Laser Spectroscopy, Shanxi University, Taiyuan 030006, China}
\author{Zhihao Xu}
\email{xuzhihao@sxu.edu.cn}
\affiliation{Institute of Theoretical Physics and State Key Laboratory of Quantum Optics Technologies and Devices, Shanxi University, Taiyuan 030006, China}

\date{\today}

\begin{abstract}
How localization emerges in open wave systems can differ fundamentally from the familiar picture in closed materials. Here we show that a quasiperiodic chain with spatially disordered non-reciprocal couplings supports a form of localization in which the relevant boundary does not separate spreading and trapped waves, but two different kinds of confinement. One regime shows erratic skin accumulation at disorder-selected positions inside the lattice, whereas the other shows conventional Anderson localization with centers distributed across the system. Adding longer-range hopping turns this transition into an anomalous mobility edge between these two localized sectors. We identify the transition using real-space profiles, Lyapunov exponents, spectral winding, and wave-packet evolution. For individual disorder realizations, wave packets drift in a direction fixed by spectral topology, while averaging over disorder cancels opposite drifts and restores conventional transport. These results provide experimentally accessible signatures of anomalous localization in engineered open systems.
\end{abstract}

\maketitle

\section{Introduction}

Non-Hermitian quantum systems arise in many open and driven platforms, where gain and loss \cite{Carl1998, Regensburger2012, Lee2016, Shuo2020, Franca2022, Peng2024, Takata2018, Sarkar2025, Roy2026}, dissipation \cite{Kunst2018, Ramy2018, Kohei2023, Liu2022, Chimdessa2025, Simon2020, Roy2026}, or effective non-unitary evolution \cite{ZhangL2021, ShuXQ2024, LeiZT2024, Sanchez2023, Leadbeater2024} are present. Typical examples include photonics \cite{Sounas2017, LiAD2023, GaoF2022, YanQC2023, MengHY2024}, acoustics \cite{Peng2024, ZhangL2021, ZhouHT2024, GuZM2023, ZhangQC2023}, electrical circuits \cite{Islam2025, Huang2023, ZhangXW2024, WangYT2025, Zhu2023}, and engineered quantum devices \cite{Sounas2017, Reisenbauer2024, Bergholtz2021, YuT2024, ZhangBW2025, Ramos2021, KohJM2025}. Non-Hermiticity can also come from non-reciprocal couplings \cite{Kohei2023, Yago2024, Veenstra2024, Reisenbauer2024, Sounas2017, Raj2026} and continuous measurement or monitored dynamics \cite{Simon2020, XuZhihaoNJP, Lau2018, Huang2023, ZhangXW2024, ZhangWG2021, LinSX2025}. In these settings, complex energy spectra \cite{Longhi2021, Ding2022, LiangJ2025, PiJH2025, HuMY2025}, amplification or attenuation \cite{Fortin2025, Leon2022, Gupta2023, Ramos2021}, and sensitivity to boundary conditions \cite{Sonu2024, Okuma2020, Bergholtz2021, Robert2024} often lead to physics that has no Hermitian counterpart.

A central phenomenon enabled by non-Hermiticity is the non-Hermitian skin effect (NHSE): under open boundary conditions (OBCs), a macroscopic number of eigenstates accumulate near the boundaries \cite{Raj2026, Li2020, LiLH2020, Okuma2020, Lin2023, Jiang2024, Kenji2024, Huang2024, Yoshida2024, Chakradhar2025, Zhao2025, HuYM2025, GuZM2022, JinWW2025, Longhi2026, ZhongJX2026, SunYY2026, ShenRZ2025, ShenRZ20252}. This accumulation invalidates the usual bulk-boundary correspondence based on Bloch bands \cite{Xiao2020, Lin2023, YaoSY2018, WuH2025, XiaoYX2024, Borgnia12023, Borgnia2020, LeeCH2019, JiangH2023}, and motivates non-Bloch descriptions and topological characterizations in the complex-energy plane \cite{Yago2024, Veenstra2024, ZhangWG2021, Raj2026, LinSX2025, Longhi2019PRR, Xiao2020, Lin2023, Longhi2021, Bergholtz2021, Yokomizo2019, WangYP2025, ZhangJH2025, LeeKW2025}. When non-Hermiticity is combined with aperiodic order, such as quasiperiodic lattices and quasicrystals, the interplay among interference, quasiperiodic modulation, and non-reciprocity can generate rich phase diagrams, including localization transitions and changes of spectral winding \cite{Jiang2019, Longhi2019, Xu2022, ZhangYL2025, WangYSX2025}.

Quasiperiodic models provide a route to localization physics without fully random disorder. The Aubry-Andr\'{e}-Harper (AAH) model is a paradigmatic example, featuring a sharp delocalization-localization transition tuned by the strength of an incommensurate onsite modulation \cite{Rossignolo2019, Cookmeyer2020, YiTC2025, LiH2023, QiR2023}. When longer-range hopping is included, generalized AAH models support mobility edges, namely energy-dependent boundaries that separate localized and extended states \cite{Ramakumar2014, Biddle2011}. These Hermitian results motivate the question of how mobility-edge physics is reshaped in non-Hermitian quasicrystals, especially when the non-Hermiticity is not spatially uniform.

Another basic issue concerns the relation between localization and dynamics. In Hermitian disordered systems, exponential localization of eigenstates is tied to the suppression of transport \cite{Weidemann2021, Sahoo2022}. However, in non-Hermitian systems this relation can be altered. Recent experiments and theory have shown that wave propagation can remain possible even when all eigenstates are exponentially localized, leading to unusual spreading that is controlled by the distribution of the imaginary parts of the eigenenergies \cite{LiB2025, Yago2024, WangYT2025, Longhi2019PRR, Balasubrahmaniyam2020, Tzortzakakis2021}. These results highlight that non-Hermitian systems may display nontrivial dynamics in regimes where a Hermitian system would be dynamically localized. This motivates a careful, phase-resolved study that combines eigenstate diagnostics, spectral topology, and real-time evolution.

A particularly interesting and less explored direction appears when an imaginary gauge field becomes spatially disordered \cite{Midya2024, Andre2000, Cheng2024, Pang2024, Sascha2021, Longhi2025PRB, DongJL2026}. In this case, non-reciprocity varies from bond to bond and cannot be described by a single uniform asymmetry. Recently, Longhi uncovered that a disordered imaginary gauge field can induce an ``erratic'' non-Hermitian skin effect (ENHSE) \cite{Longhi2025}: instead of accumulating at a boundary as in the conventional NHSE, eigenstates show macroscopic accumulation around seemingly irregular bulk positions selected by the disorder realization. This phenomenon poses a challenge for localization characterization. On the one hand, both ENHSE states and conventional Anderson-localized states are spatially confined, and their fractal dimensions can vanish in the thermodynamic limit. On the other hand, their physical origins and spectral and topological properties can be very different. It is therefore important to identify practical diagnostics that can distinguish these phases and to clarify whether a mobility edge can separate two distinct types of localized states, rather than separating extended and localized states as in the Hermitian case.

In this work, we study a one-dimensional (1D) non-Hermitian quasicrystal described by a generalized AAH chain with a Bernoulli-type disordered imaginary gauge field and a quasiperiodic onsite potential. The model includes both nearest-neighbor (NN) and next-nearest-neighbor (NNN) hoppings, allowing us to investigate within a unified framework both the standard AAH limit and the emergence of mobility-edge physics. For the standard non-Hermitian AAH limit, we identify a localization transition from a fully ENHSE phase to a fully Anderson-localized phase. This transition is anomalous in the sense that it connects two distinct confined regimes, rather than the conventional extended and localized phases of the Hermitian AAH model. We show that, although the fractal dimension vanishes in both phases, the Lyapunov exponent and the fluctuation of the eigenstate center of mass sharply distinguish them. With weak NNN hopping, we further uncover an anomalous mobility edge at the same location as in the Hermitian generalized AAH model, but with a different physical meaning in the original non-Hermitian frame: it separates Anderson-localized states from ENHSE-type macroscopic-accumulation states rather than extended states. We demonstrate that this anomalous localization structure is reflected in real-space morphology, finite-size spectral properties under periodic boundary conditions (PBCs), spectral winding, and wave-packet dynamics. Our results reveal a qualitatively distinct form of localization physics in non-Hermitian quasicrystals with disordered imaginary gauge fields and provide practical spectral, topological, and dynamical probes for its identification. 

\section{Results and Discussion}
\subsection{Model and Hamiltonian}

We consider a 1D non-Hermitian quasicrystal with a spatially disordered imaginary gauge field. This system is described by the non-Hermitian tight-binding Hamiltonian 
\begin{align}
\hat{H}=\sum_{j}(J_j^L\hat{c}_j^{\dagger}\hat{c}_{j+1}+J_j^{R}\hat{c}_{j+1}^{\dagger}\hat{c}_{j}) +\sum_{j}(\tilde{J}_j^{L}\hat{c}_j{^\dagger}\hat{c}_{j+2}+\tilde{J}_j^{R}\hat{c}_{j+2}^{\dagger}\hat{c}_j)+\sum_{j}\lambda_j\hat{c}^{\dagger}_j\hat{c}_j, \label{eq1}
\end{align}
where $\hat{c}_j$ ($\hat{c}_j^{\dagger}$) annihilates (creates) a particle at site $j$. Here $J_j^{L(R)}$ denote the left (right) NN hopping amplitudes, while $\tilde{J}_j^{L(R)}$ denote the left (right) NNN hopping amplitudes. A spatially disordered imaginary gauge field is implemented via the asymmetric hoppings $J_j^{L}=J_1\exp{(-h_j)}$, $J_j^{R}=J_1\exp{(h_j)}$, $\tilde{J}_j^{L}=J_2\exp{[-(h_j+h_{j+1})]}$, and $\tilde{J}_j^{R}=J_2\exp{(h_j+h_{j+1})}$, where $h_j$ is a Bernoulli random variable taking values $\pm \Delta h$ with equal probability. The quasiperiodic onsite potential is chosen as $\lambda_j=\lambda\cos{(2\pi\alpha j)}$, where $\lambda$ is the modulation strength and $\alpha$ is an irrational number characterizing the incommensurate structure.

In the Hermitian limit ($\Delta h=0$), the model reduces to a generalized AAH model. When the NNN hopping is switched off ($J_2=0$), one recovers the standard AAH model, which exhibits a global delocalized-localized transition at $\lambda=2J_1$. In this case, all the eigenstates are extended for $\lambda<2J_1$, while all the eigenstates become localized for $\lambda>2J_1$. When weak NNN coupling is present $J_2\ne 0$, the generalized AAH model no longer shows a single global transition. Instead, an energy-dependent mobility edge emerges, separating localized and extended states. For $J_2\ll J_1$, the mobility edge is given by \cite{Ramakumar2014, Biddle2011}
\begin{align}
E_c=\frac{1}{2}\lambda\left(\frac{J_1}{J_2}+\frac{J_2}{J_1} \right)-\frac{J_1^2}{J_2}. \label{eq2}
\end{align}
By varying $\lambda$, the system undergoes a sequence of localization transitions, from a fully extended phase to an intermediate phase with coexisting localized and extended states, and finally to a fully localized phase. 

In the non-Hermitian case ($\Delta h\ne 0$) for $\lambda=0$, the disordered imaginary gauge field can induce an ENHSE~\cite{Longhi2025}. This behavior is distinct from both the conventional NHSE, where eigenstates accumulate at one boundary, and Anderson localization, where localized states are distributed throughout the lattice without macroscopic boundary accumulation. In the ENHSE regime, eigenstates exhibit macroscopic localization around seemingly irregular positions in the bulk, typically featuring a dominant localization peak accompanied by possible satellite peaks whose locations depend sensitively on the specific disorder realization. From a characterization perspective, both NHSE and Anderson localization can display a positive Lyapunov exponent $\gamma>0$ and a vanishing fractal dimension $D=0$. In contrast, in the ENHSE regime all eigenstates display $\gamma=0$ while still having $D=0$, highlighting its unconventional nature compared with standard non-Hermitian skin and Anderson-localized phases.

In the following, we investigate anomalous localization transitions in the generalized non-Hermitian AAH model for the two cases $J_2=0$ and $J_2\ne 0$. Throughout, we set $J_1=1$ as the energy unit. Without loss of generality, we fix $\alpha=(\sqrt{5}-1)/2$ and $\Delta h=0.5$.

\subsection{Anomalous localization transitions}

\begin{figure*}[htbp]
  \centering
  \includegraphics[clip, width=0.9\columnwidth]{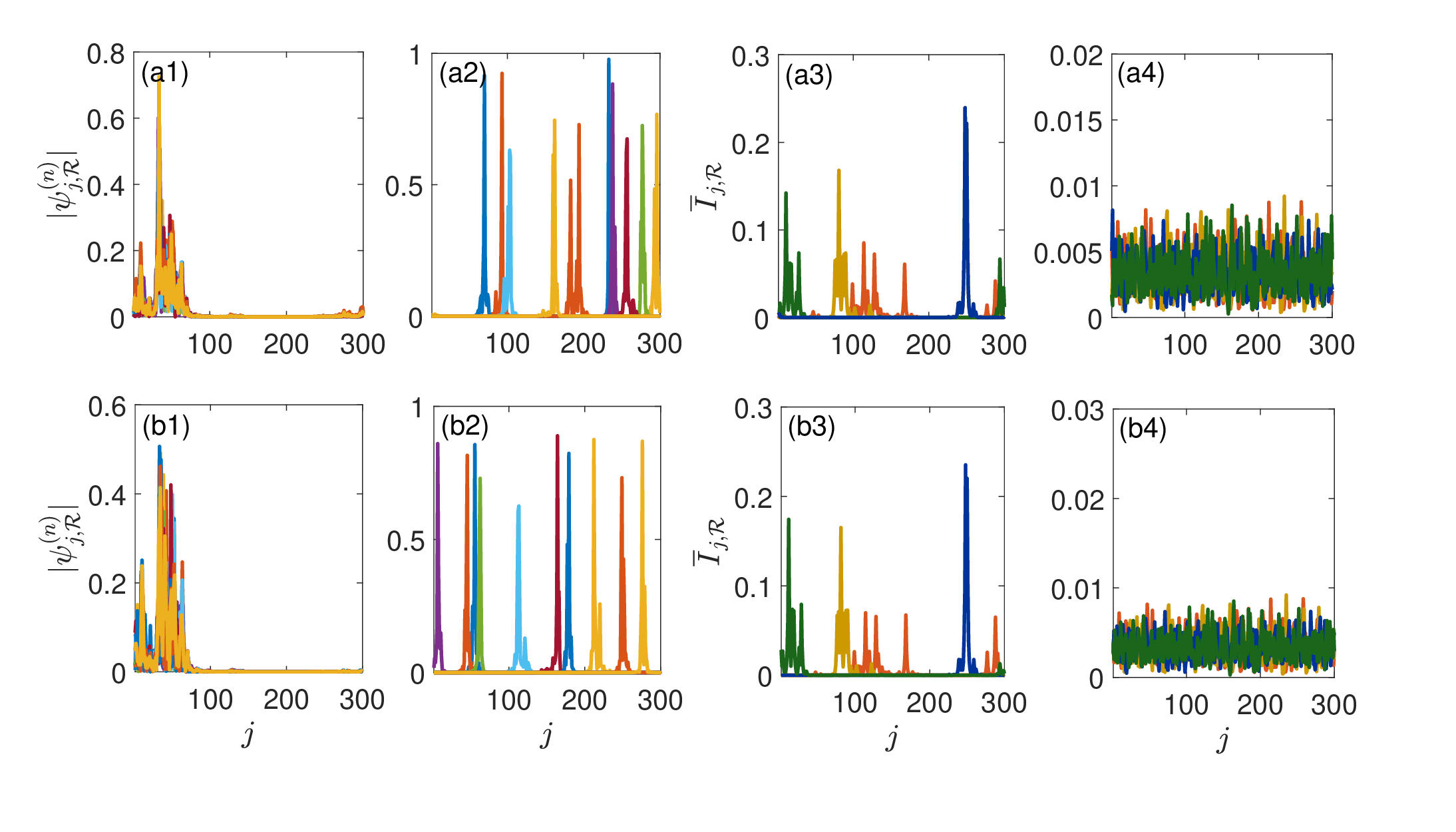}
  \caption{Representative eigenstate profiles and spatial distributions for the non-Hermitian AAH model in Eq.~\eqref{eq1}. Panels (a1), (a2), (b1), and (b2) show the amplitudes $|\psi_{j,\mathcal{R}}^{(n)}|$ of ten randomly selected eigenstates for a representative realization $\mathcal{R}$ of the imaginary gauge field under PBCs [(a1) $\lambda=1$; (a2) $\lambda=3$] and OBCs [(b1) $\lambda=1$; (b2) $\lambda=3$]. Eigenstates in the ENHSE phase exhibit macroscopic accumulation around disorder-selected bulk sites under both PBCs and OBCs, whereas in the localized phase their localization centers are distributed throughout the lattice. Panels (a3), (a4), (b3), and (b4) show $\overline{I}_{j,\mathcal{R}}$ for four distinct disorder realizations under PBCs [(a3) $\lambda=1$; (a4) $\lambda=3$] and OBCs [(b3) $\lambda=1$; (b4) $\lambda=3$]. The accumulation position varies from one realization to another in the ENHSE phase, whereas the eigenstate-averaged profile becomes nearly featureless in the localized phase. Here, $J_1=1$, $J_2=0$, $\Delta h=0.5$, and $N=300$.}
  \label{Fig1}
\end{figure*}

We begin with the non-Hermitian AAH model in Eq.~\eqref{eq1} with $J_2=0$, which exhibits an anomalous localization transition from a fully ENHSE phase to a fully localized phase. Figures~\ref{Fig1}(a1), \ref{Fig1}(a2), \ref{Fig1}(b1), and \ref{Fig1}(b2) show the profiles $|\psi_{j,\mathcal{R}}^{(n)}|$ of ten randomly selected eigenstates for a representative realization $\mathcal{R}$ of the imaginary gauge field at different values of $\lambda$ under different boundary conditions. For weak quasiperiodic modulation [$\lambda=1$; Figs.~\ref{Fig1}(a1) and \ref{Fig1}(b1)], these eigenstates exhibit macroscopic accumulation around seemingly irregular bulk positions under both PBCs and OBCs. Typically, each eigenstate displays a dominant peak accompanied by several weaker satellite peaks, which is characteristic of the fully ENHSE phase. By contrast, for strong modulation [$\lambda=3$; Figs.~\ref{Fig1}(a2) and \ref{Fig1}(b2)], the eigenstates become strongly localized, with localization centers distributed throughout the lattice under both boundary conditions, indicating a conventional fully localized phase. Notably, the spatial structures of the eigenstates in the presence of the Bernoulli-type imaginary gauge field are largely insensitive to the boundary conditions: both the macroscopic accumulation pattern in the fully ENHSE regime and the distribution of localization centers in the fully localized regime remain qualitatively similar under PBCs and OBCs.

To further illustrate the disorder dependence of the eigenstate distribution, we also examine the quantity
\begin{equation}
\overline{I}_{j,\mathcal{R}}=\frac{1}{N}\sum_{n=1}^{N}|\psi_{j,\mathcal{R}}^{(n)}|^2, \label{eqI}
\end{equation}
which represents the spatial distribution averaged over all eigenstates for a given disorder realization $\mathcal{R}$ \cite{Longhi2025}. Unless otherwise specified, all spatial diagnostics are computed from normalized right eigenstates of the non-Hermitian Hamiltonian. Figures~\ref{Fig1}(a3), \ref{Fig1}(a4), \ref{Fig1}(b3), and \ref{Fig1}(b4) show $\overline{I}_{j,\mathcal{R}}$ for four distinct disorder realizations under both PBCs and OBCs. For weak quasiperiodic modulation [$\lambda=1$; Figs.~\ref{Fig1}(a3) and \ref{Fig1}(b3)], $\overline{I}_{j,\mathcal{R}}$ exhibits pronounced macroscopic accumulation at different lattice positions for different disorder realizations, reflecting the realization-dependent nature of the ENHSE under both boundary conditions. By contrast, for strong modulation [$\lambda=3$; Figs.~\ref{Fig1}(a4) and \ref{Fig1}(b4)], $\overline{I}_{j,\mathcal{R}}$ becomes nearly uniform across the lattice and depends only weakly on the disorder realization, consistent with the conventional fully localized phase. Motivated by this qualitative similarity between PBCs and OBCs, we focus on PBCs in the following discussion for clarity.

\begin{figure*}[htbp]
  \centering
  \includegraphics[clip, width=0.7\columnwidth]{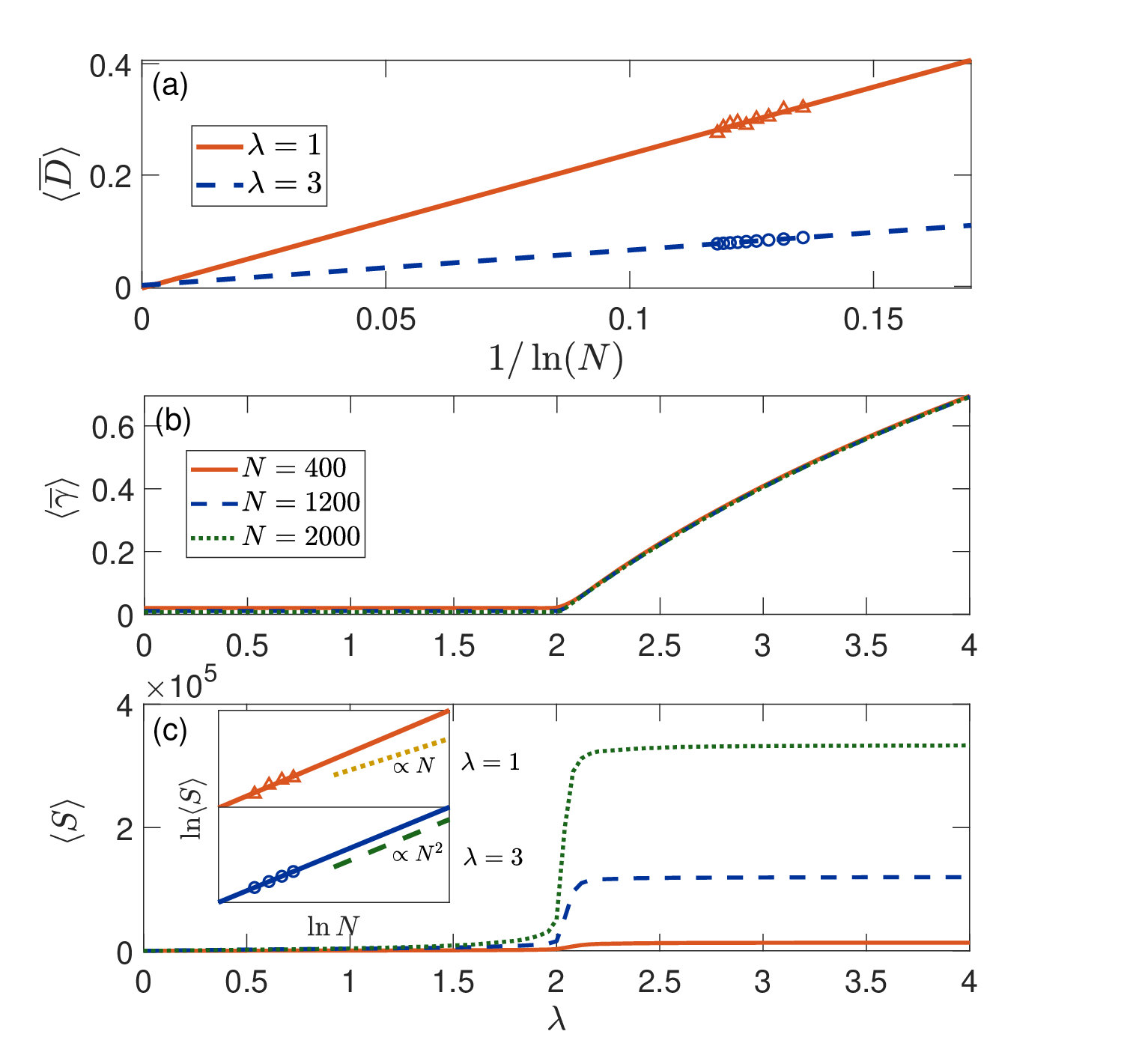} 
  \caption{Diagnostics of the anomalous localization transition. (a) Disorder-averaged mean fractal dimension $\langle \overline{D} \rangle$ versus $1/\ln N$ for $\lambda=1$ and $\lambda=3$, showing $\langle \overline{D} \rangle\to 0$ in the thermodynamic limit for both phases, confirming that both phases are spatially confined. (b) Disorder-averaged mean Lyapunov exponent $\langle \overline{\gamma} \rangle$ as a function of $\lambda$ for different system sizes, indicating a transition at $\lambda_c=2$. (c) Disorder-averaged center-of-mass fluctuation $\langle S \rangle$ versus $\lambda$ for different system sizes, with a pronounced jump near $\lambda=2$, indicating a transition in the eigenstate distribution. Inset: log-log plot of $\langle S \rangle$ versus system size $N$ for $\lambda=1$ and $\lambda=3$, demonstrating that the center-of-mass fluctuation scales as $O(N)$ and $O(N^2)$, respectively. Here, $J_1=1$, $J_2=0$, $\Delta h=0.5$, and the disorder average is taken over $N_{\mathcal{R}}=100$ realizations of the imaginary gauge field.}
  \label{Fig2}
\end{figure*}

To quantitatively distinguish the fully ENHSE phase from the fully localized phase, we examine both the fractal dimension and the Lyapunov exponent. For a given realization $\mathcal{R}$ of the imaginary gauge field, the spatial fractal property of an eigenstate $\psi^{(n)}$ with eigenvalue $E_n$ is characterized by the fractal dimension
\begin{equation}
D_{\mathcal{R}}^{(n)}=-\lim_{N\to\infty}\left[\frac{\ln \mathrm{IPR}_{\mathcal{R}}^{(n)}}{\ln N}\right],
\label{eq3}
\end{equation}
where the inverse participation ratio is defined as $\mathrm{IPR}_{\mathcal{R}}^{(n)}=\sum_{j=1}^{N}|\psi_{j,\mathcal{R}}^{(n)}|^4$, and $N$ is the system size. According to Ref.~\cite{Longhi2025}, $D_{\mathcal{R}}^{(n)}$ is expected to vanish in the thermodynamic limit in both the fully ENHSE and fully localized phases. Consistent with this expectation, Fig.~\ref{Fig2}(a) shows the disorder-averaged mean fractal dimension $\langle \overline{D} \rangle=N_{\mathcal{R}}^{-1}\sum_{\mathcal{R}=1}^{N_{\mathcal{R}}}\overline{D}_{\mathcal{R}}$, where $\overline{D}_{\mathcal{R}}=(1/N)\sum_{n=1}^{N}D_{\mathcal{R}}^{(n)}$ and $N_{\mathcal{R}}$ is the number of realizations of the imaginary gauge field, as a function of $1/\ln N$ for $\lambda=1$ and $\lambda=3$, respectively. In both cases, $\langle \overline{D} \rangle$ decreases approximately linearly with decreasing $1/\ln N$ and extrapolates to zero in the thermodynamic limit. Both ENHSE states and Anderson-localized states are spatially confined in the thermodynamic limit. Although ENHSE states show macroscopic accumulation at disorder-selected bulk positions rather than conventional exponentially localized peaks spread throughout the lattice, their inverse participation ratio still scales in such a way that the fractal dimension vanishes in the thermodynamic limit. Therefore, the fractal dimension alone cannot identify the transition. We therefore turn to the Lyapunov exponent. For a given realization $\mathcal{R}$ and each eigenstate, we define
\begin{equation}
\gamma_{\mathcal{R}}^{(n)}=\left|\lim_{N\to\infty}\frac{1}{N}\ln\left|\frac{\psi_{N,\mathcal{R}}^{(n)}}{\psi_{1,\mathcal{R}}^{(n)}}\right|\right|,
\label{eq4}
\end{equation}
which characterizes the asymptotic exponential growth or decay of the wave-function amplitude. Although Eq.~\eqref{eq4} is written in terms of asymptotic wave-function amplitudes, its validity in the present model follows from the non-unitary gauge mapping to the Hermitian generalized AAH chain, as detailed in Methods. We denote the corresponding mean value by $\langle \overline{\gamma} \rangle=N_{\mathcal{R}}^{-1}\sum_{\mathcal{R}=1}^{N_{\mathcal{R}}}\overline{\gamma}_{\mathcal{R}}$, where $\overline{\gamma}_{\mathcal{R}}=(1/N)\sum_{n=1}^{N}\gamma_{\mathcal{R}}^{(n)}$. As detailed in Methods, the Lyapunov exponent undergoes a transition from zero to a finite value at $\lambda=2$ in the thermodynamic limit. Figure~\ref{Fig2}(b) shows $\langle \overline{\gamma} \rangle$ as a function of the quasiperiodic strength $\lambda$ for different system sizes. One finds that $\langle \overline{\gamma} \rangle \to 0$ for $\lambda<2$, whereas it becomes finite for $\lambda>2$ and increases with $\lambda$, signaling the onset of exponential localization. Moreover, $\langle \overline{\gamma} \rangle$ is essentially insensitive to the system size. Combining these diagnostics, we identify a transition from the fully ENHSE phase to the fully localized phase at $\lambda_c=2$. Although $\langle \overline{D} \rangle \to 0$ in both regimes, $\langle \overline{\gamma} \rangle$ changes from vanishing for $\lambda<2$ to finite for $\lambda>2$, thus providing a sharp criterion for the transition.

Furthermore, motivated by the distinct wave-function profiles shown in Fig.~\ref{Fig1}, we introduce an additional diagnostic based on the fluctuation of the eigenstate center of mass. We define
\begin{equation}
S_{\mathcal{R}}=\frac{1}{N}\sum_{n=1}^{N}\left(x_{\mathcal{R}}^{(n)}-\bar{x}_{\mathcal{R}}\right)^2,
\label{eq5}
\end{equation}
where
\begin{equation}
x_{\mathcal{R}}^{(n)}=\frac{\sum_{j=1}^{N} j|\psi_{j,\mathcal{R}}^{(n)}|^2}{\sum_{j=1}^{N}|\psi_{j,\mathcal{R}}^{(n)}|^2}
\label{eq6}
\end{equation}
is the center-of-mass position of the $n$th eigenstate, and
$\bar{x}_{\mathcal{R}}=(1/N)\sum_{n=1}^{N}x_{\mathcal{R}}^{(n)}$
is the corresponding mean center-of-mass position for the disorder realization $\mathcal{R}$. Figure~\ref{Fig2}(c) shows the disorder-averaged fluctuation
$\langle S \rangle=N_{\mathcal{R}}^{-1}\sum_{\mathcal{R}=1}^{N_{\mathcal{R}}}S_{\mathcal{R}}$
as a function of $\lambda$ for different system sizes. In the fully ENHSE phase, for a fixed realization of the imaginary gauge field, many eigenstates exhibit similar skin-accumulated profiles and therefore have center-of-mass positions clustered around a few disorder-selected bulk locations. By contrast, in the fully localized phase, although each eigenstate is individually localized, the localization centers of different eigenstates are distributed approximately uniformly throughout the lattice. As a result, $\langle S \rangle$ exhibits a pronounced jump at $\lambda=2$, consistent with the transition identified from $\langle \overline{\gamma} \rangle$. The inset of Fig.~\ref{Fig2}(c) further presents a log-log plot of $\langle S \rangle$ versus system size for $\lambda=1$ and $\lambda=3$. The fitted slopes are close to $1$ and $2$, respectively, demonstrating that in the fully ENHSE phase the center-of-mass fluctuation scales linearly with system size, $\langle S \rangle \propto N$, whereas in the fully localized phase it scales quadratically, $\langle S \rangle \propto N^2$. This difference reflects the distinct distributions of eigenstate centers of mass in the two phases: in the ENHSE regime they remain relatively clustered around a few preferred bulk positions, while in the Anderson-localized regime they spread across the entire chain. The different scaling of $\langle S \rangle$ therefore provides an additional and intuitive diagnostic for distinguishing the two phases.

\begin{figure*}[htbp]
  \centering
  \includegraphics[clip, width=0.7\columnwidth]{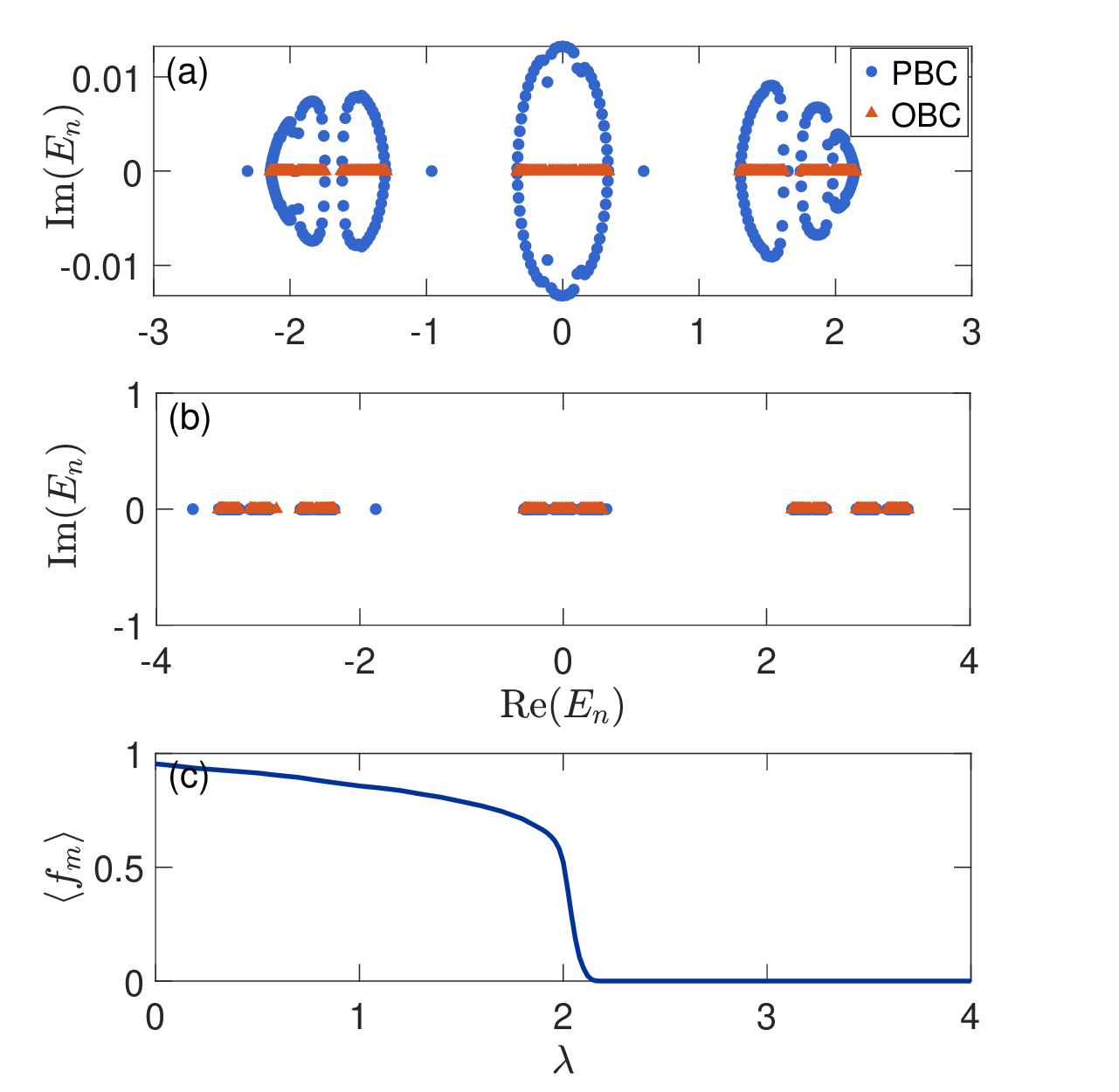}
  \caption{Spectral signatures of the anomalous localization transition. Energy spectra are shown for a representative realization $\mathcal{R}$ of the imaginary gauge field under PBCs and OBCs at (a) $\lambda=1$ and (b) $\lambda=3$, with $N=300$. For this generic finite-size realization, the spectra under OBCs are entirely real in both phases, while the PBC spectrum changes from complex to real as the system enters the localized phase. (c) Disorder-averaged fraction of complex-energy eigenstates, $\langle f_m \rangle$, versus $\lambda$ under PBCs for $N=300$, obtained with $N_{\mathcal{R}}=100$, showing a complex-to-real spectral transition near $\lambda=2$. Parameters: $J_1=1$, $J_2=0$, and $\Delta h=0.5$.}
  \label{Fig3}
\end{figure*}

Many non-Hermitian extensions of the AAH model have been extensively investigated, with particular emphasis on the interplay between the complex-to-real spectral transition and the localization transition. Figures~\ref{Fig3}(a) and \ref{Fig3}(b) show the energy spectra for a finite system of size $N=300$ under different boundary conditions at $\lambda=1$ and $\lambda=3$, respectively, for a representative realization $\mathcal{R}$ of the imaginary gauge field. For $\lambda=1$ in the fully ENHSE phase [Fig.~\ref{Fig3}(a)], the spectrum $E_n^{\mathrm{PBC}}$ is generally complex under PBCs, whereas $E_n^{\mathrm{OBC}}$ becomes purely real under OBCs. By contrast, in the fully localized phase [Fig.~\ref{Fig3}(b)], the spectra under both PBCs and OBCs are entirely real. To quantify the spectral transition, we define the disorder-averaged fraction of complex-energy eigenstates as $\langle f_m \rangle = N_{\mathcal{R}}^{-1}\sum_{\mathcal{R}=1}^{N_{\mathcal{R}}} f_{m,\mathcal{R}}$, where $f_{m,\mathcal{R}}=N_{\mathcal{R}}^{(\mathrm{im})}/N$, and $N_{\mathcal{R}}^{(\mathrm{im})}$ denotes the number of eigenenergies with nonzero imaginary parts for the realization $\mathcal{R}$. Figure~\ref{Fig3}(c) shows $\langle f_m \rangle$ as a function of $\lambda$ under PBCs for $N=300$. At $\lambda=0$, one finds $\langle f_m \rangle \approx 1$, and it decreases gradually for $\lambda<2$. For $\lambda>2$, $\langle f_m \rangle$ drops to zero, with a sharp decrease occurring around $\lambda=2$, indicating a transition from a complex spectrum to a purely real one. This complex-to-real spectral transition occurs at the same critical point as the anomalous localization transition identified above, revealing a direct correspondence between spectral reality and the localization properties of the system. It is worth noting that, for a finite system size, the spatial average of the imaginary gauge field for a given realization $\mathcal{R}$, $\overline{h}_{\mathcal{R}}=(1/N)\sum_{j=1}^{N}h_{\mathcal{R},j}$ generally deviates from zero. In the ENHSE regime, this finite $\overline{h}_{\mathcal{R}}$ introduces a residual non-Hermiticity under PBCs and can therefore produce nonzero imaginary parts in the PBC spectrum. As shown in Supplementary Note 1 and Supplementary Figure 1, this effect diminishes with increasing system size: the typical magnitude of $\overline{h}_{\mathcal{R}}$ vanishes as $N\to\infty$, and correspondingly the maximum imaginary part of the PBC eigenenergies also tends to zero in the thermodynamic limit. Moreover, as shown in Supplementary Note 2 and Supplementary Figure 2, for finite systems one may also consider realizations satisfying $\overline{h}_{\mathcal{R}}=0$. In such cases, the PBC spectrum is already entirely real even in the ENHSE regime, and the finite-size-induced spectral complexification disappears. This further confirms that the complex PBC spectrum observed at finite $N$ is not an intrinsic feature of the ENHSE phase, but rather a consequence of the residual non-Hermiticity associated with finite $\overline{h}_{\mathcal{R}}$.

\begin{figure*}[htbp]
  \centering
  \includegraphics[clip, width=1\columnwidth]{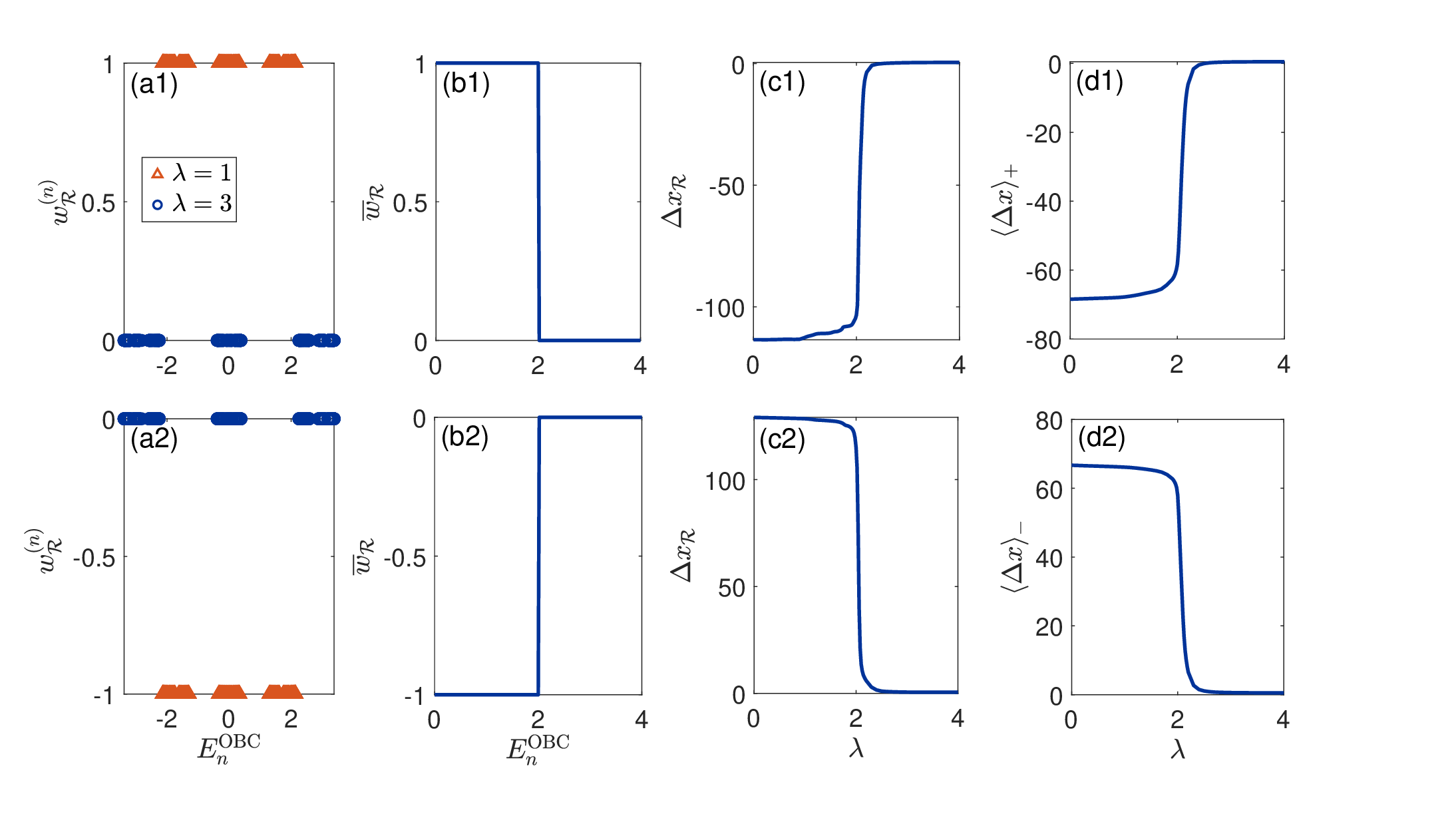}
  \caption{Correlation between spectral winding and real-space accumulation. (a1) $w_{\mathcal{R}}^{(n)}$ versus $E_n^{\mathrm{OBC}}$ at $\lambda=1$ and $3$ for a representative realization $\mathcal{R}$ with $\overline{w}_{\mathcal{R}}=+1$ in the ENHSE phase. (b1) $\overline{w}_{\mathcal{R}}$ versus $\lambda$ for the same realization. (c1) $\Delta x_{\mathcal{R}}$ versus $\lambda$ for the same realization. (d1) Realization-averaged center-of-mass shift $\langle \Delta x\rangle_{+}$ versus $\lambda$ for realizations with $\overline{w}_{\mathcal{R}}=+1$. (a2) Same as (a1), but for another representative realization with $\overline{w}_{\mathcal{R}}=-1$ in the ENHSE phase. (b2) $\overline{w}_{\mathcal{R}}$ versus $\lambda$ for the same realization. (c2) $\Delta x_{\mathcal{R}}$ versus $\lambda$ for the same realization. (d2) Realization-averaged center-of-mass shift $\langle \Delta x\rangle_{-}$ versus $\lambda$ for realizations with $\overline{w}_{\mathcal{R}}=-1$. In (d1) and (d2), the averages are performed separately over the two subsets classified by the sign of $\overline{w}_{\mathcal{R}}$, with $N_{\mathcal{R}}^{(\pm)}=100$. The two representative realizations shown in panels (a1)–(d1) and (a2)–(d2) differ only in the Bernoulli sequence of the imaginary gauge field. These results indicate that both the winding number and the center-of-mass undergo a transition at $\lambda_c=2$. The sign of the winding number depends on the specific realization and correlates with the direction of the center-of-mass shift. Parameters: $J_1=1$, $J_2=0$, $\Delta h=0.5$, and $N=300$.} \label{Fig4}
\end{figure*}

Unlike the Hermitian case, our non-Hermitian AAH model can exhibit a spectral topological transition in the complex-energy plane at finite size \cite{WangYSX2025, Xu2022, Longhi2019}. To characterize the spectral topology, we introduce the winding number $w$, defined as~\cite{Xu2022, Yokomizo2019, Longhi2025PRB}
\begin{equation}
w=\int_{0}^{2\pi}\frac{\mathrm{d}\varphi}{2\pi i}\,\partial_{\varphi}\ln\det\left[\hat{H}(\varphi)-E_{B}\right],
\label{eq7}
\end{equation}
where $\hat{H}(\varphi)$ is the Hamiltonian in Eq.~\eqref{eq1} under PBCs with a flux $\varphi$ threaded through the ring, and $E_B$ is a reference energy. This winding number counts how many times the complex PBC spectrum encircles $E_B$ as $\varphi$ is varied from $0$ to $2\pi$. In our calculations, we take the OBC eigenenergies $E_n^{\mathrm{OBC}}$ as the reference energies and evaluate the corresponding winding numbers $w_{\mathcal{R}}^{(n)}$ with $E_B=E_n^{\mathrm{OBC}}$ for a given realization $\mathcal{R}$ of the imaginary gauge field. Figure~\ref{Fig4}(a1) shows $w_{\mathcal{R}}^{(n)}$ as a function of $E_n^{\mathrm{OBC}}$ at $\lambda=1$ and $\lambda=3$ for a representative realization $\mathcal{R}$. For $\lambda=1$ in the fully ENHSE phase, the PBC spectral loops encircle each reference energy $E_n^{\mathrm{OBC}}$, yielding a nontrivial winding number $w_{\mathcal{R}}^{(n)}=+1$. By contrast, for $\lambda=3$ in the fully localized phase, the spectrum no longer winds around any $E_n^{\mathrm{OBC}}$, and the winding number becomes trivial, $w_{\mathcal{R}}^{(n)}=0$, for all reference energies. To characterize the overall topological response, we further define the mean winding number $\overline{w}_{\mathcal{R}}$ by averaging $w_{\mathcal{R}}^{(n)}$ over all reference energies $E_n^{\mathrm{OBC}}$, and plot $\overline{w}_{\mathcal{R}}$ as a function of $\lambda$ in Fig.~\ref{Fig4}(b1). For the finite system size considered and the chosen realization $\mathcal{R}$, $\overline{w}_{\mathcal{R}}$ remains pinned at $+1$ for $\lambda<2$ and then drops abruptly to $\overline{w}_{\mathcal{R}}=0$ for $\lambda>2$, signaling a spectral topological transition. This behavior is consistent with the complex-to-real spectral transition and the localization transition discussed above. We emphasize that the sign of the nontrivial winding in the fully ENHSE phase depends on the specific realization of the imaginary gauge field. For other realizations, $w_{\mathcal{R}}^{(n)}$ may take the value $-1$ instead of $+1$, as illustrated in Fig.~\ref{Fig4}(a2) for $\lambda=1$, with the corresponding behavior of $\overline{w}_{\mathcal{R}}$ shown in Fig.~\ref{Fig4}(b2). Nevertheless, upon increasing $\lambda$ into the fully localized phase, the mean winding number always becomes trivial, namely $\overline{w}_{\mathcal{R}}=0$.

The macroscopic accumulation of the eigenstates in the ENHSE phase is also reflected in the mean center-of-mass position, whose sign depends on the realization of the imaginary gauge field. To quantify this effect, we define the center-of-mass shift $\Delta x_{\mathcal{R}}=\bar{x}_{\mathcal{R}}-(N+1)/2$, and show its behavior in Figs.~\ref{Fig4}(c1) and \ref{Fig4}(c2) for the same two representative realizations used in Figs.~\ref{Fig4}(a1), \ref{Fig4}(b1), \ref{Fig4}(a2), and \ref{Fig4}(b2). In the ENHSE phase ($\lambda<2$), $\Delta x_{\mathcal{R}}$ remains finite and its sign follows that of the mean winding number: realizations with $\overline{w}_{\mathcal{R}}=+1$ exhibit a negative shift, indicating macroscopic accumulation toward the left half of the chain, whereas realizations with $\overline{w}_{\mathcal{R}}=-1$ exhibit a positive shift, indicating accumulation toward the right half. Once the system enters the fully localized phase ($\lambda>2$), the center-of-mass shift rapidly approaches zero in both cases, consistent with the disappearance of directional accumulation and the onset of nearly uniform distribution of localization centers over the lattice. To make this correlation statistically transparent, we further group the realizations according to the sign of the mean winding number in the ENHSE phase and average within each subset separately. The resulting quantities, $\langle \Delta x\rangle_{\pm}=\left(N_{\mathcal{R}}^{(\pm)}\right)^{-1}\sum_{\mathcal{R}\in\mathcal{S}_{\pm}}\Delta x_{\mathcal{R}}$, are shown in Figs.~\ref{Fig4}(d1) and \ref{Fig4}(d2), where $\mathcal{S}_{\pm}$ denotes the set of realizations satisfying $\overline{w}_{\mathcal{R}}=\pm1$ in the ENHSE phase. One finds that $\langle \Delta x\rangle_{+}<0$ and $\langle \Delta x\rangle_{-}>0$ throughout the ENHSE regime, whereas both quantities approach zero in the fully localized phase. Therefore, after resolving the realizations by the sign of the spectral winding, the macroscopic center-of-mass shift is seen to be locked to the winding sign. This establishes a direct correlation between spectral topology and real-space accumulation in the finite-size non-Hermitian AAH model.

\subsection{Anomalous mobility edges}

\begin{figure*}[htbp]
  \centering
  \includegraphics[clip, width=0.9\columnwidth]{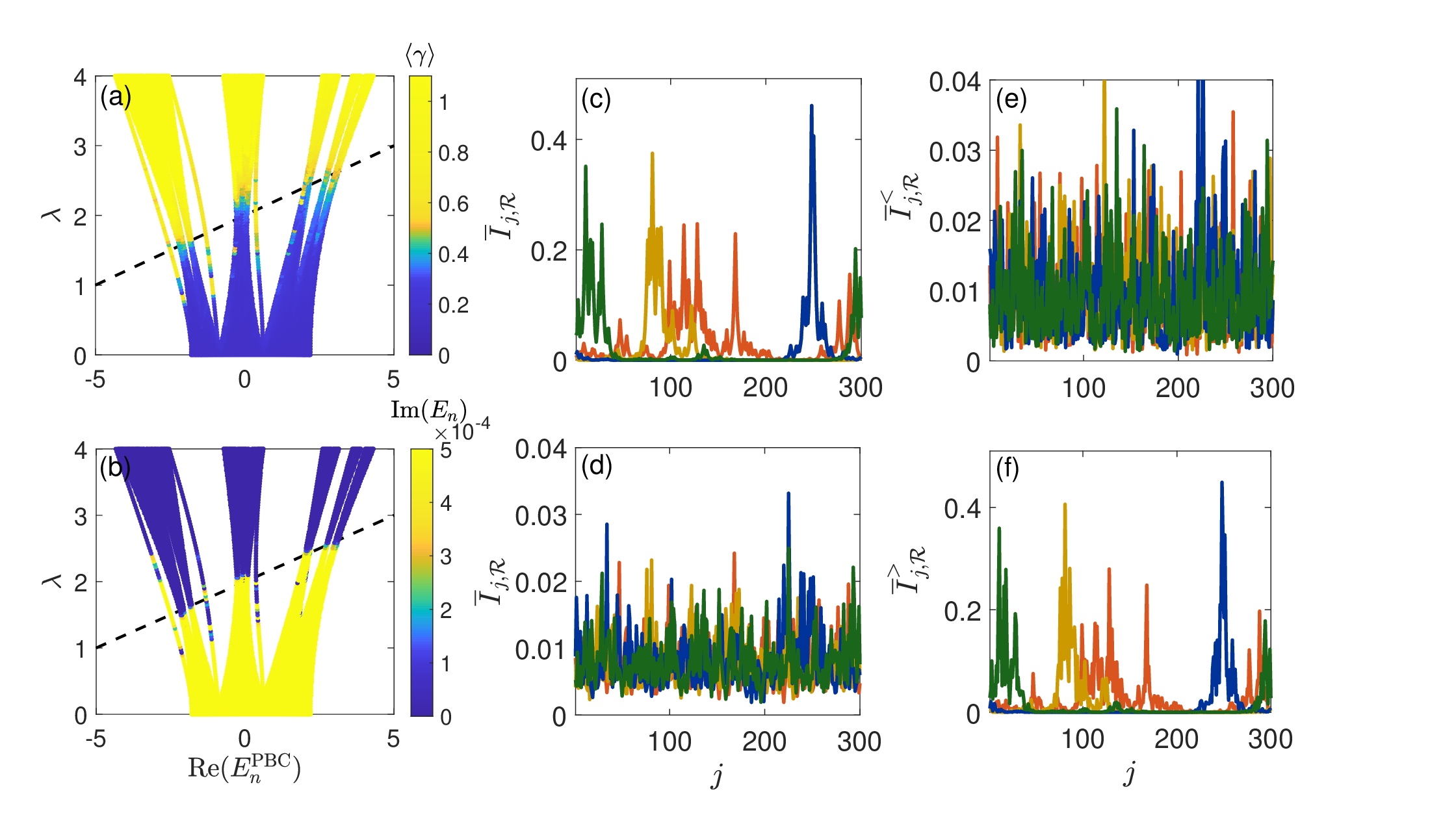}
  \caption{Anomalous mobility edge in the generalized non-Hermitian AAH model. (a) Localization phase diagram under PBCs in the $(\mathrm{Re}(E_n^{\mathrm{PBC}}),\lambda)$ plane, characterized by the disorder-averaged Lyapunov exponent $\langle \gamma \rangle$ and averaged over $N_{\mathcal{R}}=100$ realizations of the imaginary gauge field. The color map reveals three regimes: fully ENHSE ($\lambda<1.5$), coexistence ($\lambda \in [1.5,2.5]$), and fully localized ($\lambda>2.5$). The black dashed line marks the anomalous mobility edge predicted by Eq.~\eqref{eq2}. (b) Density plot of $\mathrm{Im}(E_n^{\mathrm{PBC}})$ in the $(\mathrm{Re}(E_n^{\mathrm{PBC}}),\lambda)$ plane, averaged over $N_{\mathcal{R}}=100$ realizations of the imaginary gauge field. For generic finite-size realizations, the real-complex spectral boundary tracks the anomalous mobility edge. (c) Realization-resolved averaged spatial profiles $\overline{I}_{j,\mathcal{R}}$ for four representative realizations at $\lambda=1$ in the fully ENHSE phase. (d) Realization-resolved averaged spatial profiles $\overline{I}_{j,\mathcal{R}}$ at $\lambda=3$ in the fully localized phase. (e) Subspace-averaged profiles $\overline{I}^{<}_{j,\mathcal{R}}$ for eigenstates with $\mathrm{Re}(E_n^{\mathrm{PBC}})<E_c$ at $\lambda=1.9$. (f) Subspace-averaged profiles $\overline{I}^{>}_{j,\mathcal{R}}$ for eigenstates with $\mathrm{Re}(E_n^{\mathrm{PBC}})>E_c$ at $\lambda=1.9$. Panels (c)--(f) show that the eigenstate-averaged profiles are strongly concentrated in the ENHSE-type sectors and nearly uniform in the Anderson-localized sectors. Here, $J_1=1$, $J_2=0.1$, $\Delta h=0.5$, and $N=300$.} 
  \label{Fig5}
\end{figure*}

Having established the anomalous transition in the standard non-Hermitian AAH limit ($J_2=0$), we now consider weak NNN coupling ($J_2\neq 0$ and $J_2\ll J_1$), where anomalous mobility edges emerge. In contrast to the $J_2=0$ case, which exhibits a single transition between a fully ENHSE phase and a fully localized phase, finite $J_2$ broadens the transition into an intermediate coexistence regime bounded by two critical modulation strengths. In the absence of the imaginary gauge field, the corresponding Hermitian generalized AAH model hosts a conventional mobility edge described by Eq.~\eqref{eq2}. In the present non-Hermitian model, the mobility edge remains at the same $E_c(\lambda)$ as in the Hermitian case (see Methods for details), but its physical meaning is different: it separates Anderson-localized states from ENHSE-type states with macroscopic accumulation at realization-dependent bulk positions, rather than extended and localized states. The mobility edge is therefore anomalous not because its analytical form is modified, but because in the original non-Hermitian frame it divides the spectrum into two qualitatively distinct localized sectors. As shown below, this distinction is manifested directly in the localization morphology and, for generic finite-size realizations, consistently reflected in spectral reality, winding topology, and wave-packet dynamics. Unless otherwise specified, we fix $J_2=0.1$ in the following calculations.

\begin{figure*}[htbp]
  \centering
  \includegraphics[clip, width=0.8\columnwidth]{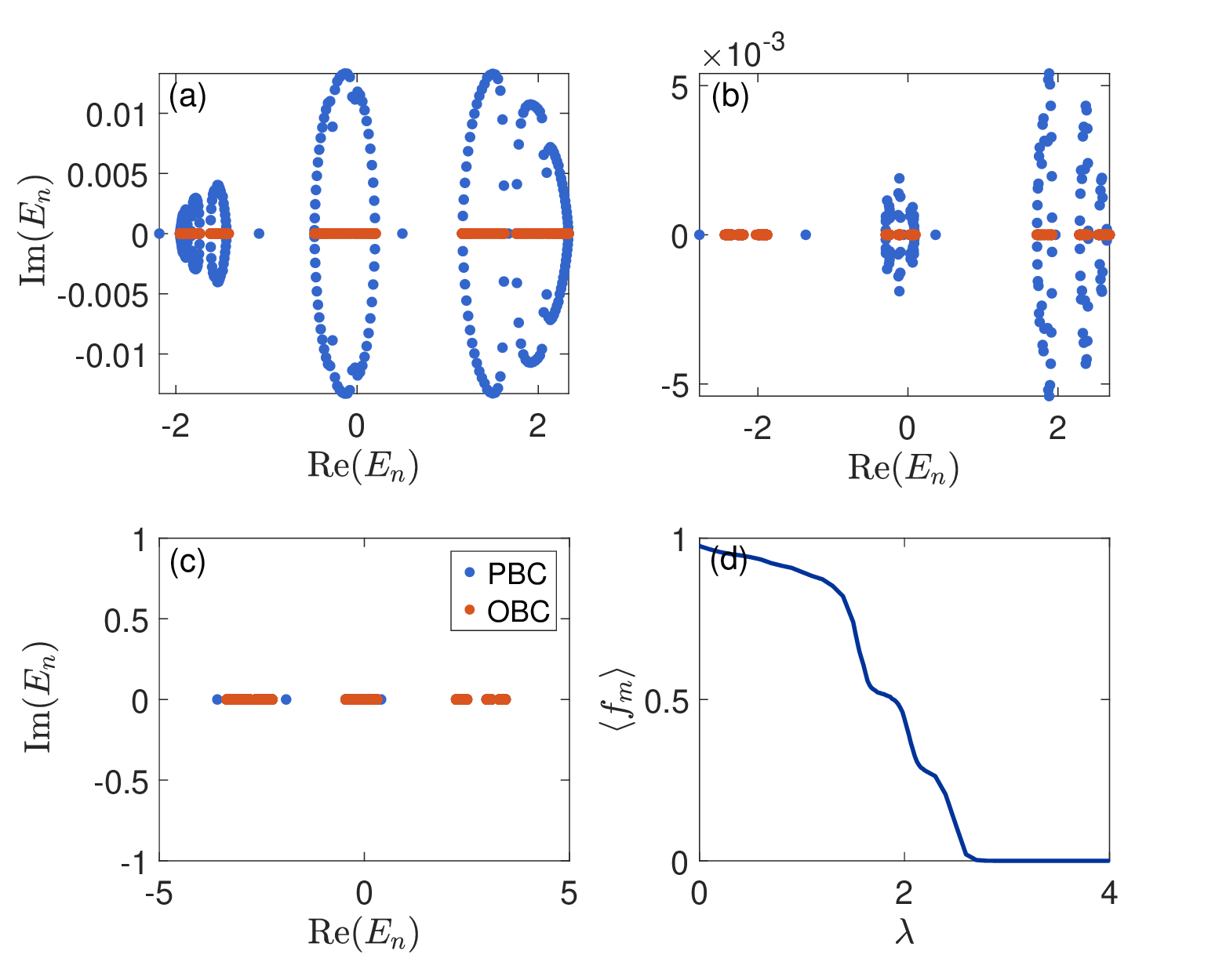}
  \caption{Finite-size spectral evolution across the anomalous mobility edge. (a)--(c) Energy spectra under PBCs and OBCs for a representative realization of the imaginary gauge field at $\lambda=1$, $1.9$, and $3$, respectively. For the representative finite-size realization, the PBC spectrum forms loops in the ENHSE phase, partially collapses to the real axis in the intermediate coexistence regime, and becomes entirely real in the localized phase. (d) Realization-averaged fraction of complex-energy eigenstates, $\langle f_m\rangle$, as a function of $\lambda$, obtained with $N_{\mathcal{R}}=100$. Here, $J_1=1$, $J_2=0.1$, $\Delta h=0.5$, and $N=300$.} 
  \label{Fig6}
\end{figure*}

Consistent with the $J_2=0$ case, we find that the fractal dimension does not provide a reliable characterization of the phase structure in the present non-Hermitian system. We therefore use the Lyapunov exponent to quantify localization and to construct the phase diagram. Figure~\ref{Fig5}(a) shows the localization phase diagram under PBCs in the $(\mathrm{Re}(E_n^{\mathrm{PBC}}),\lambda)$ plane, obtained by averaging over $N_{\mathcal{R}}=100$ realizations of the imaginary gauge field. The color scale denotes the disorder-averaged Lyapunov exponent $\langle \gamma \rangle$, and the black dashed line marks the anomalous mobility edge predicted by Eq.~\eqref{eq2}. For $\lambda<\lambda_{c1}=1.5$, the system is in the fully ENHSE phase. As illustrated in Fig.~\ref{Fig5}(c) for $\lambda=1$, the realization-resolved averaged profiles $\overline{I}_{j,\mathcal{R}}$ for four representative disorder realizations exhibit pronounced macroscopic accumulation around irregular bulk positions, accompanied by weaker satellite peaks. For $\lambda>\lambda_{c2}=2.5$, the system enters the fully localized phase, in which the eigenstates remain localized while their localization centers are distributed nearly uniformly along the chain. As shown in Fig.~\ref{Fig5}(d) for $\lambda=3$, the corresponding profiles $\overline{I}_{j,\mathcal{R}}$ become nearly uniform, reflecting the broad distribution of localization centers across the lattice. In the intermediate regime $\lambda_{c1}<\lambda<\lambda_{c2}$, an anomalous mobility edge emerges in the $(\mathrm{Re}(E_n^{\mathrm{PBC}}),\lambda)$ plane and separates Anderson-localized states from ENHSE-type states. Taking $\lambda=1.9$ as an example, the mobility edge is located at $E_c\simeq -0.4$. To visualize the two sectors separated by the mobility edge, we further introduce the realization-resolved profiles averaged over the subsets of eigenstates satisfying $\mathrm{Re}(E_n^{\mathrm{PBC}})<E_c$ and $\mathrm{Re}(E_n^{\mathrm{PBC}})>E_c$, respectively, $\overline{I}^{<}_{j,\mathcal{R}}=N_{<}^{-1}\sum_{n\in\mathcal{S}_{<}}\left|\psi_{j,\mathcal{R}}^{(n)}\right|^2$ and $\overline{I}^{>}_{j,\mathcal{R}}=N_{>}^{-1}\sum_{n\in\mathcal{S}_{>}}\left|\psi_{j,\mathcal{R}}^{(n)}\right|^2$, where $\mathcal{S}_{<}$ and $\mathcal{S}_{>}$ denote the sets of eigenstates with $\mathrm{Re}(E_n^{\mathrm{PBC}})<E_c$ and $\mathrm{Re}(E_n^{\mathrm{PBC}})>E_c$, and $N_{<}$ and $N_{>}$ are the corresponding numbers of eigenstates. As shown in Fig.~\ref{Fig5}(e), eigenstates with $\mathrm{Re}(E_n^{\mathrm{PBC}})<E_c$ exhibit conventional Anderson-localized features, with $\overline{I}^{<}_{j,\mathcal{R}}$ showing an almost uniform spatial distribution. By contrast, states with $\mathrm{Re}(E_n^{\mathrm{PBC}})>E_c$ display ENHSE-type macroscopic accumulation under different realizations of the imaginary gauge field, as reflected by the strongly inhomogeneous profile $\overline{I}^{>}_{j,\mathcal{R}}$ in Fig.~\ref{Fig5}(f).

For generic finite-size realizations, the generalized non-Hermitian AAH model with weak NNN coupling exhibits a real-complex spectral crossover that correlates with the anomalous mobility edge. To verify its correspondence with the localization transition, we compute the imaginary part of the PBC eigenenergies, $\mathrm{Im}(E_n^{\mathrm{PBC}})$, as a function of $\mathrm{Re}(E_n^{\mathrm{PBC}})$ and $\lambda$, as shown in Fig.~\ref{Fig5}(b), obtained by averaging over $N_{\mathcal{R}}=100$ realizations of the imaginary gauge field. One can clearly identify two critical modulation strengths, $\lambda_{c1}$ and $\lambda_{c2}$, which coincide with those of the localization transition. For generic finite-size realizations with nonzero $\overline h_{\mathcal R}$, most or all PBC eigenenergies are complex for $\lambda<\lambda_{c1}$, indicating that the system is entirely in the ENHSE regime. For $\lambda>\lambda_{c2}$, all eigenenergies become purely real, corresponding to the fully localized phase. In the intermediate regime $\lambda_{c1}<\lambda<\lambda_{c2}$, an energy boundary $E_c(\lambda)$ emerges and tracks the anomalous mobility edge. Specifically, eigenenergies with $\mathrm{Re}(E_n^{\mathrm{PBC}})<E_c(\lambda)$ are purely real, whereas those with $\mathrm{Re}(E_n^{\mathrm{PBC}})>E_c(\lambda)$ remain complex. These results demonstrate that the real-complex spectral transition is fully consistent with the localization transition in the generalized non-Hermitian AAH model. To further elucidate the interplay between the anomalous mobility edge and spectral topology, we next examine how the energy spectrum evolves under different boundary conditions as $\lambda$ is varied. Figures~\ref{Fig6}(a)-\ref{Fig6}(c) show the spectra for a representative realization of the imaginary gauge field at $\lambda=1$, $1.9$, and $3$, respectively. In the fully ENHSE regime ($\lambda<\lambda_{c1}$), the spectrum splits into well-separated clusters. Under PBCs, the eigenvalues are complex and form loop structures in the complex-energy plane, as shown in Fig.~\ref{Fig6}(a) for $\lambda=1$. As $\lambda$ increases into the intermediate regime, e.g., at $\lambda=1.9$, the spectrum becomes energy selective across the mobility edge: under PBCs, eigenvalues with $\mathrm{Re}(E_n^{\mathrm{PBC}})<E_c$ collapse onto the real axis, whereas those with $\mathrm{Re}(E_n^{\mathrm{PBC}})>E_c$ remain complex and continue to form loops in the complex-energy plane, as shown in Fig.~\ref{Fig6}(b). Upon further increasing $\lambda$ into the fully localized regime ($\lambda>\lambda_{c2}$), the entire PBC spectrum becomes purely real, as shown in Fig.~\ref{Fig6}(c) for $\lambda=3$. By contrast, under OBCs the spectrum $E_n^{\mathrm{OBC}}$ remains purely real throughout the parameter range considered. To quantify the real-complex spectral transition under PBCs, we plot in Fig.~\ref{Fig6}(d) the realization-averaged fraction of complex-energy eigenstates, $\langle f_m\rangle$, as a function of $\lambda$. For $\lambda<\lambda_{c1}$, $\langle f_m\rangle$ remains close to unity, indicating that almost all eigenenergies are complex. In the intermediate regime $\lambda_{c1}<\lambda<\lambda_{c2}$, $\langle f_m\rangle$ decreases continuously from 1 to 0, reflecting the gradual shrinking of the complex-energy sector as the mobility edge moves through the spectrum. For $\lambda>\lambda_{c2}$, $\langle f_m\rangle\to 0$, indicating that the PBC spectrum has become entirely real. This behavior further confirms the consistency between the real-complex spectral transition and the anomalous localization transition.

\begin{figure*}[htbp]
  \centering
  \includegraphics[clip, width=0.8\columnwidth]{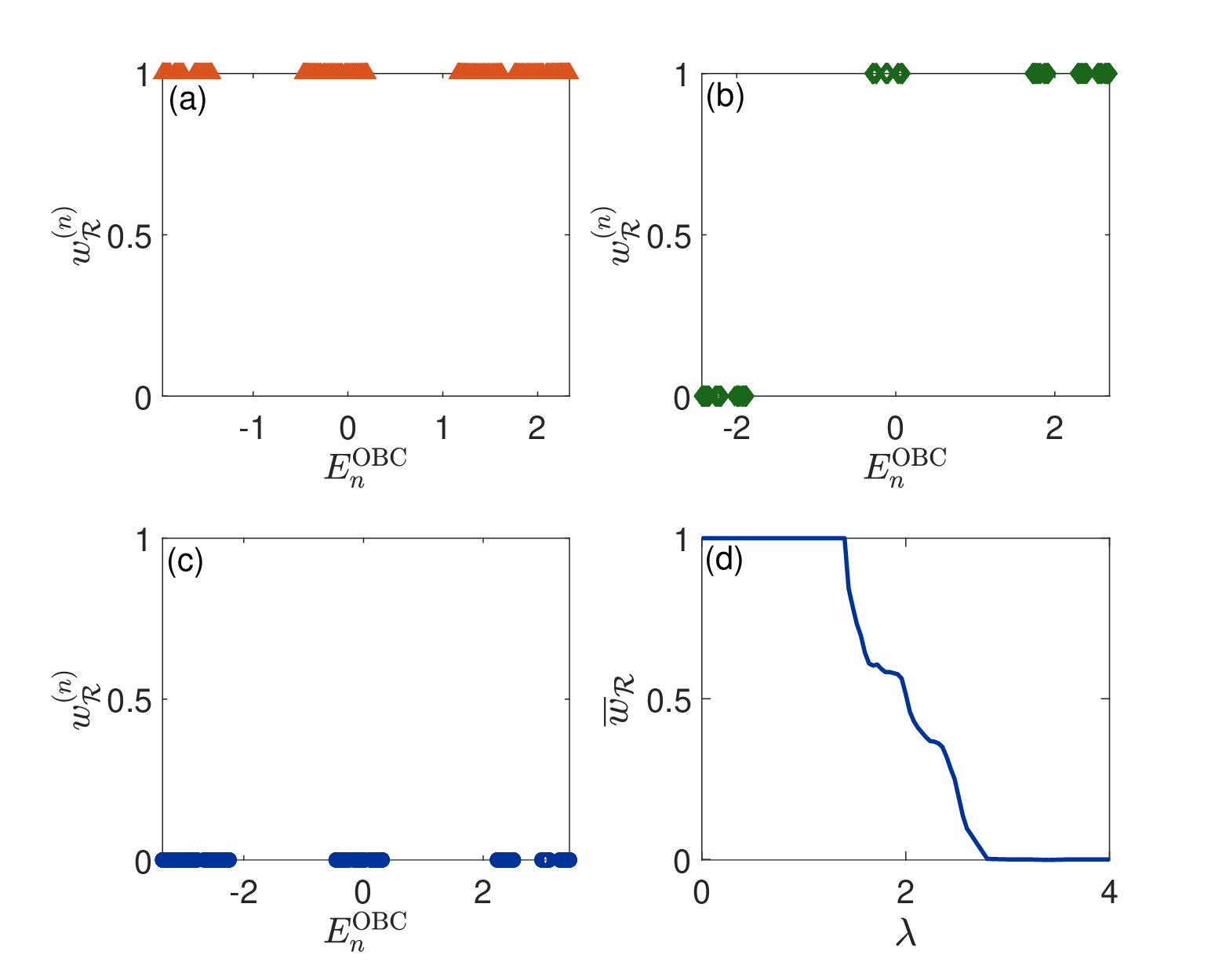}
  \caption{Topological characterization of the anomalous mobility edge. (a)--(c) Winding number $w_{\mathcal{R}}^{(n)}$, evaluated by taking the OBC eigenenergies $E_n^{\mathrm{OBC}}$ as reference energies, for $\lambda=1$, $1.9$, and $3$, respectively. (d) Mean winding number $\overline{w}_{\mathcal{R}}$ as a function of $\lambda$ for the same realization $\mathcal{R}$ of the imaginary gauge field. The winding number undergoes an energy-dependent transition associated with the anomalous mobility edge, consistent with the finite-size spectral transition. The realization used in panels (a)-(d) is the same as that in Figs.~\ref{Fig6}(a)--\ref{Fig6}(c). Here, $J_1=1$, $J_2=0.1$, $\Delta h=0.5$, and $N=300$.} 
  \label{Fig7}
\end{figure*}

\begin{figure*}[htbp]
  \centering
  \includegraphics[clip, width=0.7\columnwidth]{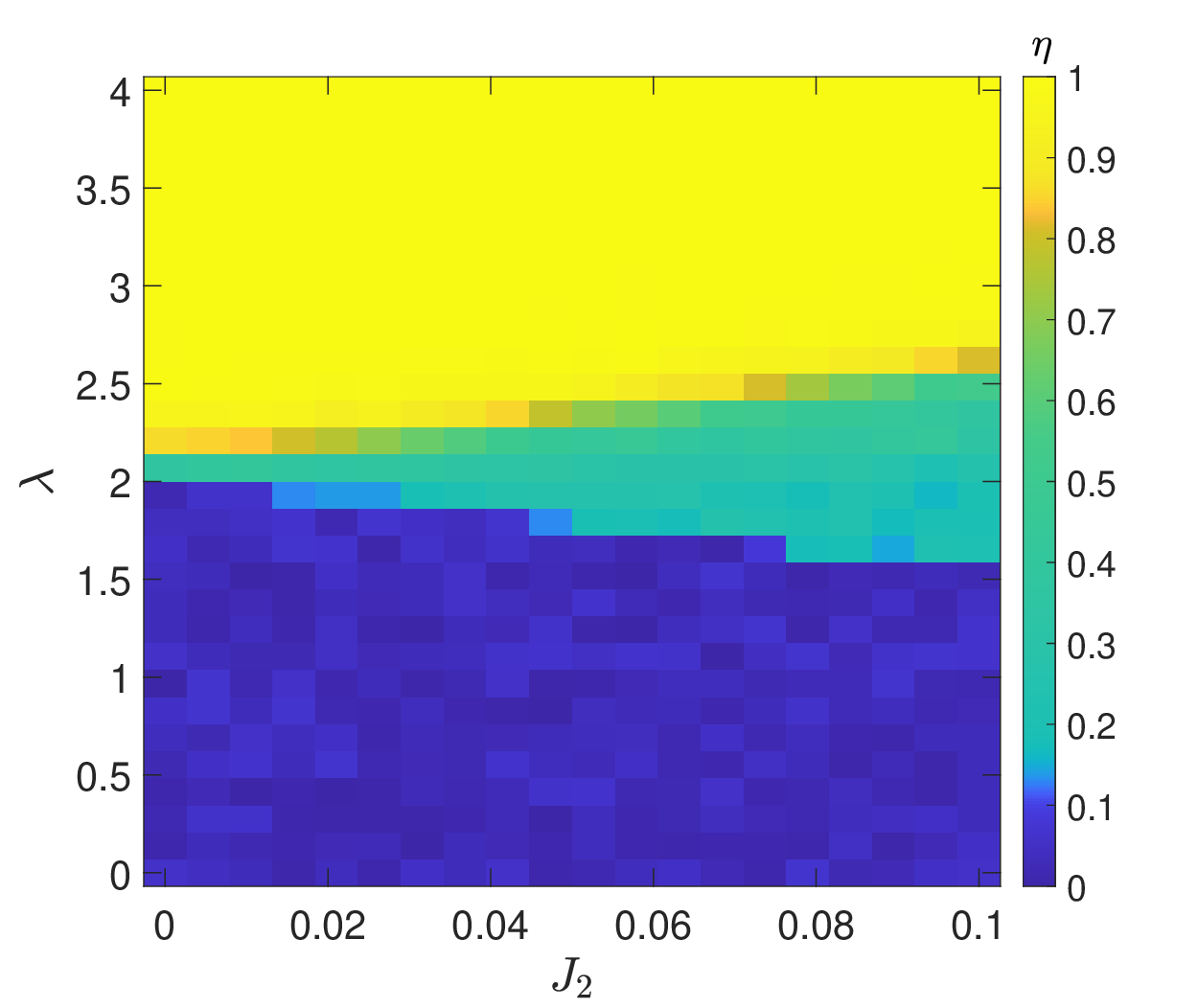}
 \caption{Global localization phase diagram. The localization phase diagram in the $(\lambda,J_2)$ plane under PBCs is characterized by the effective fractal exponent $\eta$ of the averaged spatial profile. The results are obtained by averaging over $N_{\mathcal{R}}=100$ realizations of the imaginary gauge field. Values of $\eta$ close to $0$ correspond to the fully ENHSE phase, values close to $1$ correspond to the fully localized phase, and intermediate values $0<\eta<1$ indicate the coexistence regime. The coexistence regime broadens monotonically as $J_2$ increases, indicating the emergence of the anomalous mobility edge. Here, $J_1=1$, $\Delta h=0.5$, and $N=300$.} 
 \label{Fig8}
\end{figure*}

To directly connect the anomalous mobility edge with spectral topology, we compute the winding number $w_{\mathcal{R}}^{(n)}$ for a given realization $\mathcal{R}$ of the imaginary gauge field by taking the OBC eigenenergies $E_n^{\mathrm{OBC}}$ as reference points for different values of $\lambda$, as shown in Figs.~\ref{Fig7}(a)-\ref{Fig7}(c). The realization $\mathcal{R}$ is chosen to be the same as that used in Figs.~\ref{Fig6}(a)-\ref{Fig6}(c). In the fully ENHSE regime, the PBC spectrum encircles each reference energy $E_n^{\mathrm{OBC}}$, yielding $w_{\mathcal{R}}^{(n)}=1$ [Fig.~\ref{Fig7}(a) for $\lambda=1$], consistent with the loop structures in the complex-energy plane. As $\lambda$ increases into the intermediate regime, the winding number becomes energy dependent. For $\lambda=1.9$ in Fig.~\ref{Fig7}(b), reference energies associated with states satisfying $\mathrm{Re}(E_n^{\mathrm{PBC}})<E_c$ are no longer enclosed by the PBC loops and therefore give $w_{\mathcal{R}}^{(n)}=0$, whereas those with $\mathrm{Re}(E_n^{\mathrm{PBC}})>E_c$ remain enclosed and retain $w_{\mathcal{R}}^{(n)}=1$. Upon further increasing $\lambda$ into the fully localized regime [Fig.~\ref{Fig7}(c) for $\lambda=3$], the PBC spectrum collapses onto the real axis, and the winding becomes trivial, with $w_{\mathcal{R}}^{(n)}=0$ for all reference energies. To characterize the overall topological response, we plot the mean winding number $\overline{w}_{\mathcal{R}}$ as a function of $\lambda$ for the same realization $\mathcal{R}$. For this representative realization, $\overline{w}_{\mathcal{R}}=1$ for $\lambda<\lambda_{c1}$ and $\overline{w}_{\mathcal{R}}=0$ for $\lambda>\lambda_{c2}$. In the intermediate regime $\lambda_{c1}<\lambda<\lambda_{c2}$, $\overline{w}_{\mathcal{R}}$ decreases continuously from 1 to 0. This behavior is fully consistent with the real-complex spectral transition and the localization transition in the presence of weak NNN hopping. Notably, changing the realization of the imaginary gauge field may flip the sign of the winding number in the ENHSE regime, whereas the energy-dependent onset of trivial winding across $E_c(\lambda)$ remains unchanged. This provides a direct topological characterization of the anomalous mobility edge.

To obtain a global view of the effect of weak NNN hopping, we map out the localization phase diagram in the $(\lambda,J_2)$ plane in Fig.~\ref{Fig8}. To distinguish the fully ENHSE phase, the fully localized phase, and the intermediate coexistence regime, we introduce an effective fractal exponent based on the realization-resolved averaged spatial profile $\overline{I}_{j,\mathcal{R}}$ defined in Eq.~\eqref{eqI}, which is different from the eigenstate fractal dimension defined in Eq.~\eqref{eq3}. We first define $\xi_{\mathcal{R}}=\sum_{j=1}^{N}\left(\overline{I}_{j,\mathcal{R}}\right)^2$, $\langle \xi \rangle=N_{\mathcal{R}}^{-1}\sum_{\mathcal{R}=1}^{N_{\mathcal{R}}}\xi_{\mathcal{R}}$, and then introduce
\begin{equation}
  \eta=-\frac{\ln \langle \xi \rangle}{\ln N}.
\end{equation}
The physical meaning of $\eta$ follows directly from the spatial structure of the averaged profile $\overline{I}_{j,\mathcal{R}}$. In the fully ENHSE phase, for a fixed disorder realization, most eigenstates exhibit macroscopic accumulation around only a few disorder-selected bulk positions. As a result, $\overline{I}_{j,\mathcal{R}}$ remains strongly concentrated on a limited set of sites, so that $\xi_{\mathcal{R}}$ stays finite and $\eta$ approaches $0$. By contrast, in the fully localized phase, although each eigenstate is individually localized, the localization centers of different eigenstates are distributed nearly uniformly over the entire chain. In this case, the averaged profile becomes approximately uniform, i.e. $\overline{I}_{j,\mathcal{R}}\sim 1/N$, which implies $\xi_{\mathcal{R}}\sim 1/N$, and therefore $\eta$ approaches $1$. In the intermediate coexistence regime, $\overline{I}_{j,\mathcal{R}}$ is neither sharply concentrated as in the fully ENHSE phase nor fully uniform as in the fully localized phase. Accordingly, $\xi_{\mathcal{R}}$ takes an intermediate value, leading to $0<\eta<1$. The effective fractal exponent $\eta$ therefore provides a convenient indicator for distinguishing the three regimes. Figure~\ref{Fig8} shows the resulting localization phase diagram under PBCs, characterized by $\eta$ and obtained by averaging over $N_{\mathcal{R}}=100$ realizations of the imaginary gauge field. For $J_2=0$, the coexistence regime is essentially absent, consistent with the single transition discussed in the previous section. Once $J_2$ becomes finite, however, an intermediate regime emerges and gradually broadens with increasing $J_2$. This indicates that weak NNN hopping converts the single anomalous transition into a finite coexistence window in which ENHSE-type and Anderson-localized states coexist in different parts of the spectrum. We have also verified that these features are insensitive to the specific choice of system size. As shown in Supplementary Note 3 and Supplementary Figure 3, the localization phase diagrams obtained for Fibonacci system sizes remain quantitatively consistent with those presented in the main text, with the transition points and the anomalous mobility edge unchanged within numerical resolution.

Our results show that introducing weak NNN hopping ($J_2\neq 0$) into the non-Hermitian AAH model gives rise to an anomalous mobility edge. Although its location $E_c(\lambda)$ is still determined by the Hermitian criterion in Eq.~\eqref{eq2}, it now separates Anderson-localized states from ENHSE-type states rather than from extended states. This interpretation is consistently supported by the Lyapunov-exponent-based localization analysis and, for generic finite-size realizations, by the energy-selective real-complex spectral transition under PBCs and the spectral winding that becomes trivial across $E_c(\lambda)$ and vanishes completely in the fully localized regime.

\subsection{Dynamical detection}

\begin{figure*}[htbp]
  \centering
  \includegraphics[clip, width=0.9\columnwidth]{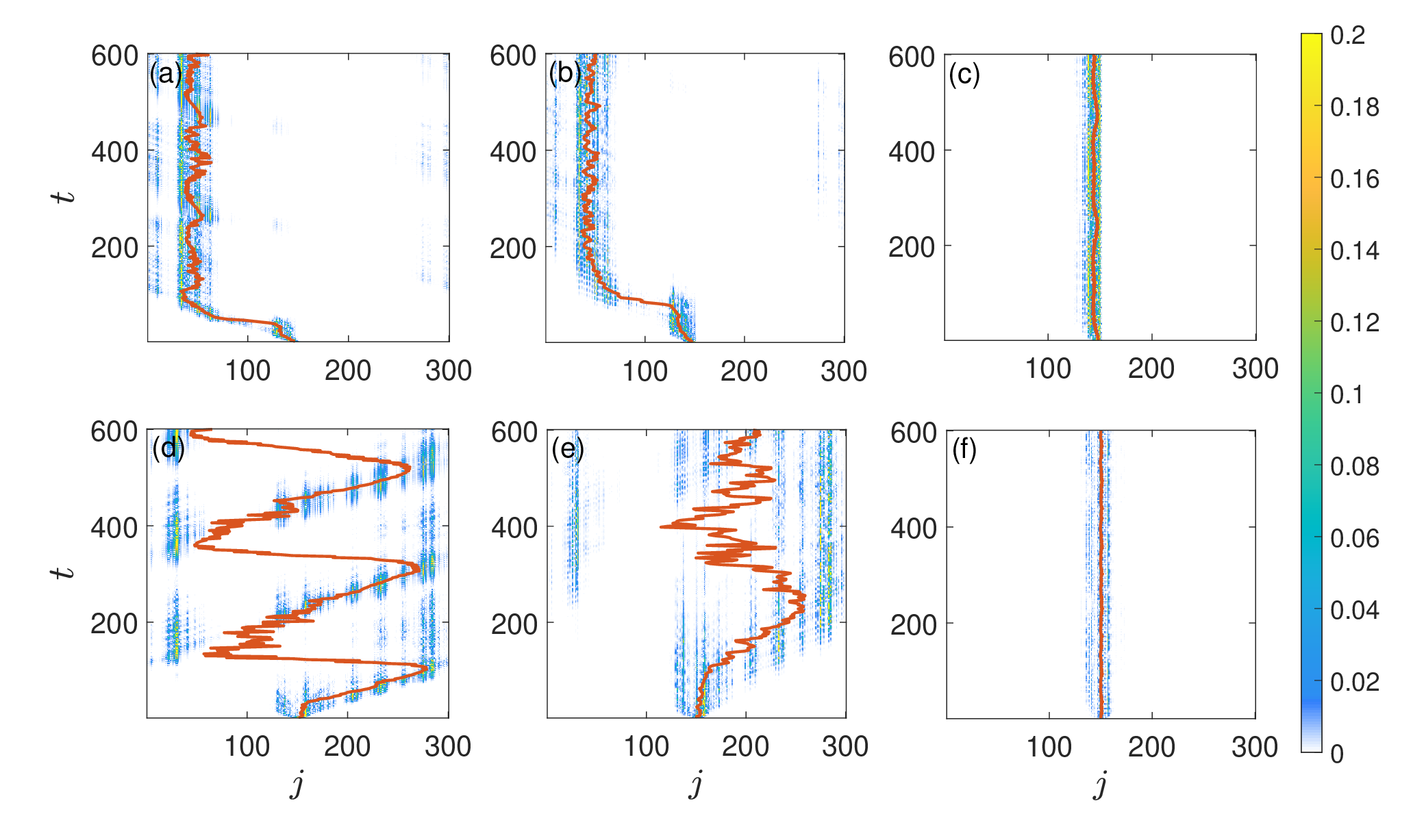}
  \caption{Realization-resolved wave-packet dynamics. Time evolution of the wave-packet density profile $\rho_{j,\mathcal{R}}(t)$ under PBCs is shown for two representative realizations $\mathcal{R}$ of the imaginary gauge field, with the initial state $|\Psi(0)\rangle=|j_0\rangle$ and $j_0=N/2$ for even $N$. The red solid line in each panel denotes the wave-packet center of mass $x_{\mathcal{R}}(t)$. Panels (a)–(c) correspond to the same realization as in Fig.~\ref{Fig7}(a)–\ref{Fig7}(c), at $\lambda=1$, $1.9$, and $3$, respectively. Panels (d)–(f) show another realization at the same values of $\lambda$, chosen such that in the fully ENHSE regime the winding numbers satisfy $w_{\mathcal{R}}^{(n)}=-1$ for all reference energies. Notably, in the ENHSE regime, the short-time drift direction depends on the sign of the spectral winding number, while the long-time trajectory is highly sensitive to the specific disorder configuration. In the fully localized regime, the wave packet remains pinned near the initial position. Here, $J_1=1$, $J_2=0.1$, $\Delta h=0.5$, and $N=300$.} 
  \label{Fig9}
\end{figure*}

To provide an intuitive picture of the dynamics, we examine the wave-packet density profile $\rho_{j,\mathcal{R}}(t)=\langle \Psi_{\mathcal{R}}(t)|j\rangle\langle j|\Psi_{\mathcal{R}}(t)\rangle=|\Psi_{j,\mathcal{R}}(t)|^2$, where $\rho_{j,\mathcal{R}}(t)$ denotes the density evolution for realization $\mathcal{R}$ of the imaginary gauge field, as shown in Fig.~\ref{Fig9}. Figures~\ref{Fig9}(a)-\ref{Fig9}(c) show the results under PBCs for the same representative realization used in Figs.~\ref{Fig7}(a)-\ref{Fig7}(c), at $\lambda=1$, $1.9$, and $3$, respectively. Figures~\ref{Fig9}(d)-\ref{Fig9}(f) show the corresponding results for another realization at the same values of $\lambda$, chosen such that in the fully ENHSE regime all reference energies $E_n^{\mathrm{OBC}}$ satisfy $w_{\mathcal{R}}^{(n)}=-1$, while in the intermediate regime the states with $\mathrm{Re}(E_n^{\mathrm{PBC}})>E_c$ also have $w_{\mathcal{R}}^{(n)}=-1$. The solid lines in each panel denote the wave-packet center of mass, $x_{\mathcal{R}}(t)=\sum_j j\,\rho_{j,\mathcal{R}}(t)$, which directly characterizes the transient drift and the long-time evolution.

For $\lambda=1$ in Fig.~\ref{Fig9}(a), the system is in the fully ENHSE phase, where all reference energies $E_n^{\mathrm{OBC}}$ have winding number $w_{\mathcal{R}}^{(n)}=1$. The initially localized wave packet quickly develops a pronounced leftward drift and evolves toward the ENHSE-favored accumulation region, leading to a corresponding leftward shift of the center of mass $x_{\mathcal{R}}(t)$. At long times, the wave function becomes pinned at realization-dependent accumulation sites, and $x_{\mathcal{R}}(t)$ remains stabilized on the left side of the chain. In the intermediate regime [Fig.~\ref{Fig9}(b), $\lambda=1.9$], where the winding is energy dependent [Fig.~\ref{Fig7}(b)], the wave packet first exhibits a short-time transient near the initial position $j_0$ and then develops a leftward drift induced by the sector with $w_{\mathcal{R}}^{(n)}=1$. Consequently, $x_{\mathcal{R}}(t)$ shows an initial plateau near $j_0$, followed by a clear shift toward the left accumulation region. By contrast, in the fully localized phase [Fig.~\ref{Fig9}(c), $\lambda=3$], all winding numbers become trivial, $w_{\mathcal{R}}^{(n)}=0$, and the wave packet remains localized near the initial site without any systematic drift, so that $x_{\mathcal{R}}(t)$ stays close to $j_0$ throughout the evolution. For the alternative realization shown in Figs.~\ref{Fig9}(d)-\ref{Fig9}(f), the fully ENHSE case at $\lambda=1$ has $w_{\mathcal{R}}^{(n)}=-1$ for all reference energies. Accordingly, the wave packet exhibits a pronounced rightward drift, and the center of mass shifts to the right. When the wave packet reaches the right boundary, PBCs wrap it around the ring, producing a rapid leftward jump of $x_{\mathcal{R}}(t)$ followed by renewed rightward drift. This process repeats and gives rise to the characteristic sawtooth trajectory in Fig.~\ref{Fig9}(d). In the intermediate regime [Fig.~\ref{Fig9}(e), $\lambda=1.9$], a similar behavior appears after a short transient near $j_0$: the wave packet develops a rightward bias, repeatedly crosses the boundary under PBCs, and recurrently accumulates at the ENHSE-favored sites. In the fully localized phase [Fig.~\ref{Fig9}(f), $\lambda=3$], the dynamics again becomes trivial, with the wave packet remaining localized near the initial site and showing no directional drift. Overall, these examples show that for a fixed realization of the imaginary gauge field, the short-time dynamics is controlled primarily by the sign of the spectral winding, while the long-time evolution is strongly realization dependent. In particular, realizations with $w_{\mathcal{R}}^{(n)}=1$ and $w_{\mathcal{R}}^{(n)}=-1$ exhibit opposite drift directions, whereas in the fully localized regime the directional drift disappears.

\begin{figure*}[htbp]
  \centering
  \includegraphics[clip, width=0.9\columnwidth]{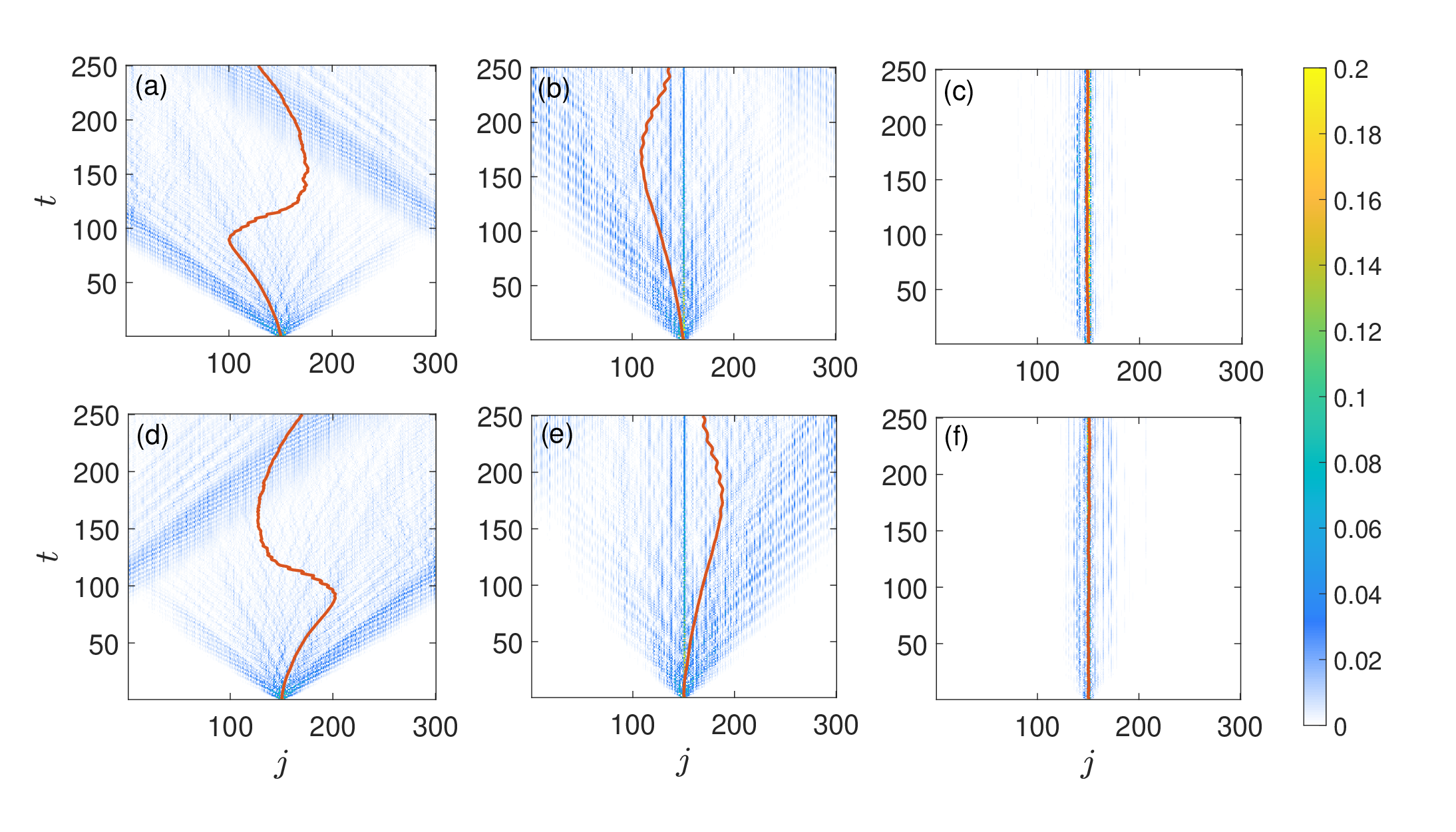}
  \caption{Winding-resolved averaged wave-packet dynamics. Group-resolved averaged wave-packet dynamics $\langle \rho_j(t)\rangle_{\pm}$ under PBCs is shown for the imaginary gauge field. The red solid line in each panel denotes the corresponding averaged center of mass $\langle x(t)\rangle_{\pm}$. Panels (a)–(c) are obtained by averaging over the subset of realizations with mean winding number $\overline{w}_{\mathcal{R}}=+1$ at $\lambda=1$, $1.9$, and $3$, respectively. Panels (d)–(f) show the corresponding results for the subset with $\overline{w}_{\mathcal{R}}=-1$ at the same values of $\lambda$. The stability of the directional drift with disorder averaging confirms a robust correlation between the sign of the winding number and the direction of non-Hermitian transport. Here, $J_1=1$, $J_2=0.1$, $\Delta h=0.5$, $N=300$, and $j_0=150$. All data are obtained by averaging over $N_{\mathcal{R}}^{(\pm)}=1000$ disorder realizations of the imaginary gauge field.} 
  \label{Fig10}
\end{figure*}

While Fig.~\ref{Fig9} shows that the detailed long-time dynamics depends sensitively on the realization of the imaginary gauge field, the short-time response already exhibits a robust realization-resolved feature: the direction of the center-of-mass motion is locked to the sign of the spectral winding. To make this relation explicit, we classify the realizations of the imaginary gauge field into two groups according to the sign of the mean winding number, $\overline{w}_{\mathcal{R}}=+1$ and $\overline{w}_{\mathcal{R}}=-1$, and perform the realization average within each group separately. In the parameter regime considered here, a fixed realization cannot host coexisting nontrivial winding sectors with opposite signs; rather, all nonzero winding numbers share the same sign within that realization, so this classification is well defined. Specifically, we define $\langle \rho_j(t)\rangle_{\pm}=\left(N_{\mathcal{R}}^{(\pm)}\right)^{-1}\sum_{\mathcal{R}\in\mathcal{S}_{\pm}}\rho_{j,\mathcal{R}}(t)$ and $\langle x(t)\rangle_{\pm}=\left(N_{\mathcal{R}}^{(\pm)}\right)^{-1}\sum_{\mathcal{R}\in\mathcal{S}_{\pm}}x_{\mathcal{R}}(t)$, where $\mathcal{S}_{\pm}$ denotes the subset of realizations satisfying $\overline{w}_{\mathcal{R}}=\pm1$. The resulting group-resolved averaged dynamics under PBCs are shown in Fig.~\ref{Fig10}: panels~\ref{Fig10}(a)-\ref{Fig10}(c) correspond to the subset with $\overline{w}_{\mathcal{R}}=+1$ at $\lambda=1$, $1.9$, and $3$, respectively, while panels~\ref{Fig10}(d)-\ref{Fig10}(f) show the corresponding results for the subset with $\overline{w}_{\mathcal{R}}=-1$. For the subset with $\overline{w}_{\mathcal{R}}=+1$, the averaged center of mass $\langle x(t)\rangle_{+}$ exhibits a pronounced leftward drift in the short-time regime for both $\lambda=1$ and $\lambda=1.9$, as shown in Figs.~\ref{Fig10}(a) and \ref{Fig10}(b). At $\lambda=1$, where the system is in the fully ENHSE phase, the wave packet drifts toward the left side of the ring and, after reaching the boundary, is wrapped around by the PBC, producing a sudden jump of the center of mass before the leftward drift resumes. At $\lambda=1.9$, in the intermediate regime, the averaged dynamics still shows an initial leftward drift, together with a localized component, consistent with the presence of the anomalous mobility edge. By contrast, at $\lambda=3$ [Fig.~\ref{Fig10}(c)], where the system is fully localized and the winding becomes trivial, the wave packet remains localized near the initial position and $\langle x(t)\rangle_{+}$ stays close to $j_0$ without systematic drift. For the subset with $\overline{w}_{\mathcal{R}}=-1$, the dynamics in Figs.~\ref{Fig10}(d)-\ref{Fig10}(f) displays the opposite response. In the short-time regime, $\langle x(t)\rangle_{-}$ drifts rightward for $\lambda=1$ and $\lambda=1.9$. At $\lambda=1$, the wave packet propagates toward the right boundary and, after boundary wrapping under PBCs, the center of mass undergoes a compensating jump before continuing its rightward drift. At $\lambda=1.9$, the averaged dynamics again contains both a directional drifting component and a localized component, but with the drift direction reversed. At $\lambda=3$ [Fig.~\ref{Fig10}(f)], the dynamics becomes trivial, with the wave packet remaining localized near $j_0$ and no directional motion in the center of mass. These results show that, once the realizations are resolved by the sign of the mean winding number, the short-time drift direction is robustly locked to the spectral winding: $\langle x(t)\rangle_{+}$ moves leftward, whereas $\langle x(t)\rangle_{-}$ moves rightward.

\begin{figure*}[htbp]
  \centering
  \includegraphics[clip, width=0.7\columnwidth]{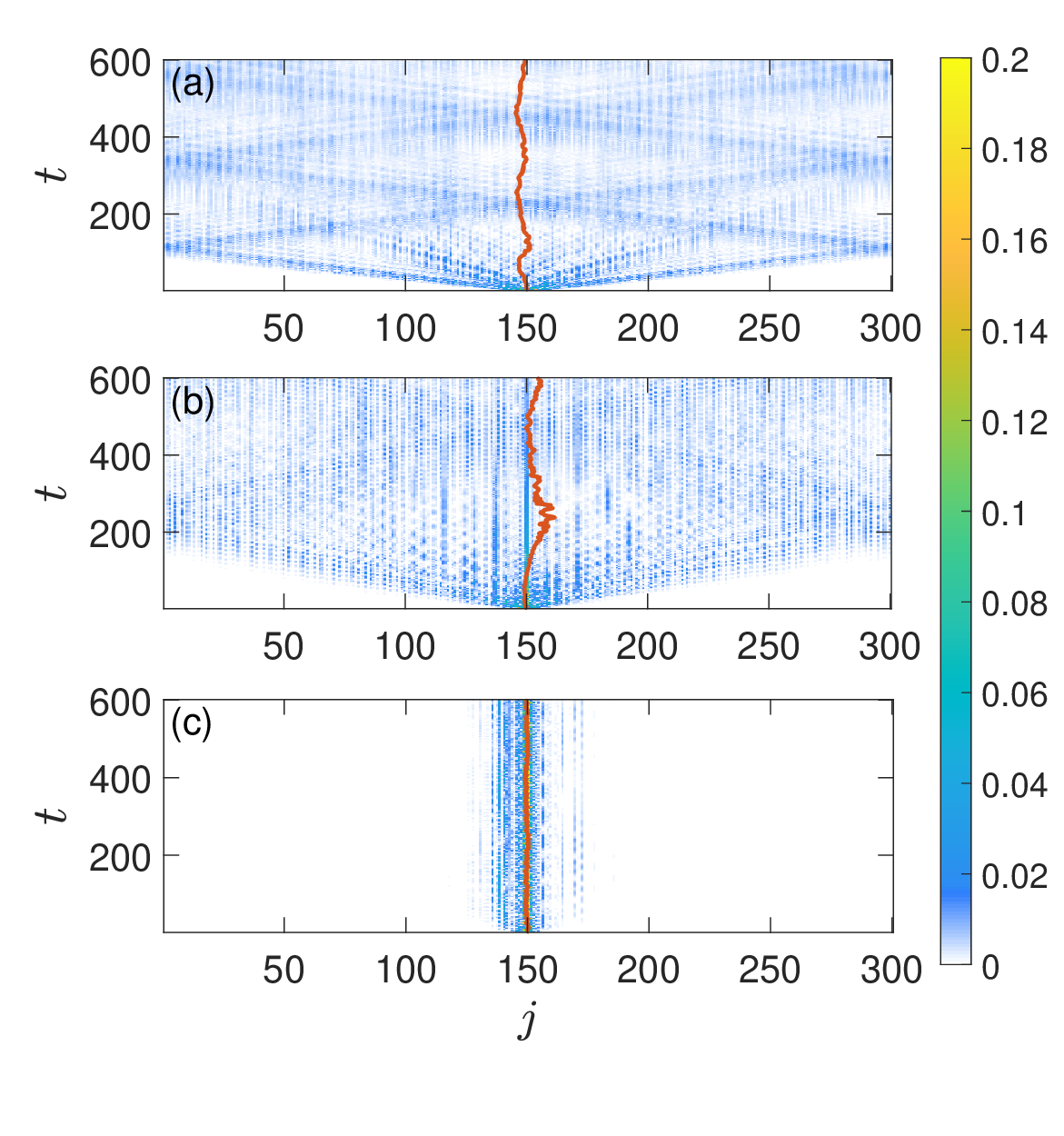}
  \caption{Disorder-averaged wave-packet dynamics. Panels (a)–(c) show the disorder-averaged density profile $\langle \rho_j(t)\rangle$ under PBCs for the imaginary gauge field at $\lambda=1$, $1.9$, and $3$, respectively, obtained by averaging over $N_{\mathcal{R}}=1000$ realizations of the imaginary gauge field. The red solid line in each panel denotes the corresponding disorder-averaged center of mass $\langle x(t)\rangle$. In contrast to Fig.~\ref{Fig10}, the directional drift vanishes with the global disorder average. Here, $J_1=1$, $J_2=0.1$, $\Delta h=0.5$, $N=300$, and $j_0=150$.} 
  \label{Fig11}
\end{figure*}

Having identified this winding-resolved dynamical structure, we finally consider the full disorder average over all realizations of the imaginary gauge field. Figures~\ref{Fig11}(a)-\ref{Fig11}(c) show the fully averaged wave-packet evolution $\langle \rho_j(t)\rangle = N_{\mathcal{R}}^{-1}\sum_{\mathcal{R}}\rho_{j,\mathcal{R}}(t)$, and the corresponding averaged center of mass $\langle x(t)\rangle = N_{\mathcal{R}}^{-1}\sum_{\mathcal{R}}x_{\mathcal{R}}(t)$, for $\lambda=1$, $1.9$, and $3$, respectively, with $N_{\mathcal{R}}=1000$. Since the two winding sectors carry opposite drift directions, their contributions largely cancel in the full disorder average. As a result, the resulting dynamics closely resembles that of the Hermitian counterpart~\cite{XuZhihaoNJP}. For $\lambda=1$, the averaged wave packet exhibits ballistic-like spreading \cite{ZhongJX2026}, while $\langle x(t)\rangle$ remains close to the initial position, reflecting the cancellation between the left- and right-drifting sectors. At $\lambda=1.9$, the averaged dynamics displays both a spreading component and a localized component near the initial site, consistent with mobility-edge-like behavior in the corresponding Hermitian model~\cite{XuZhihaoNJP}; meanwhile, the net center-of-mass drift remains strongly suppressed for the same cancellation reason. In the fully localized phase at $\lambda=3$, both $\langle \rho_j(t)\rangle$ and $\langle x(t)\rangle$ remain localized near the initial position throughout the evolution. These results show that the group-resolved average reveals the robust dynamical signature of spectral topology, whereas the full disorder average is dominated by the mutual cancellation between the two opposite winding sectors and therefore restores Hermitian-like transport behavior.

In short, for an individual realization of the imaginary gauge field, the short-time wave-packet dynamics exhibits a clear winding-dependent drift, whereas the long-time evolution remains strongly realization dependent. After resolving the realizations into two groups according to the sign of the mean winding number, the group-averaged dynamics retains the opposite directional responses of the two winding sectors. In contrast, the fully averaged dynamics over all realizations is dominated by the mutual cancellation between these opposite sectors and therefore largely restores Hermitian-like transport behavior.

\section{Conclusion}

We studied anomalous localization in a 1D non-Hermitian quasiperiodic lattice with a spatially disordered imaginary gauge field, focusing on both the standard AAH limit ($J_2=0$) and the generalized model with weak NNN hopping ($J_2\neq 0$). For $J_2=0$, we identified an anomalous transition from a fully ENHSE phase to a fully localized phase at $\lambda_c=2$. Although the fractal dimension vanishes in both regimes, the Lyapunov exponent and the fluctuation of the eigenstate center of mass sharply distinguish them. For generic finite-size realizations, this transition is further accompanied by a complex-to-real spectral change under PBCs and a winding-number change from nontrivial to trivial. With weak NNN hopping, we demonstrated an anomalous mobility edge at the same $E_c(\lambda)$ as in the Hermitian generalized AAH model, but with a distinct physical meaning in the original non-Hermitian frame: it separates Anderson-localized states from ENHSE-type macroscopic-accumulation states rather than extended states. We further showed that this anomalous localization structure is reflected in real-space morphology, spectral reality, winding topology, and wave-packet dynamics. In particular, single realizations exhibit winding-dependent drift with realization-dependent long-time behavior, winding-resolved averaging preserves the opposite directional responses of the two sectors, whereas full disorder averaging largely cancels them and restores Hermitian-like transport. Our results provide a unified non-Hermitian characterization of anomalous localization and mobility edges in quasicrystals with disordered imaginary gauge fields.

\section{Methods}

\subsection{Lyapunov exponent calculation}

For the generalized non-Hermitian AAH model, the stationary Schr\"odinger equation for an eigenstate $|\psi^{(n)}\rangle$ with energy $E_n$ is
\begin{equation}
E_n\psi_j^{(n)}=\tilde{J}_j^{L}\psi_{j+2}^{(n)} +J_j^{L}\psi_{j+1}^{(n)}+J_{j-1}^{R}\psi_{j-1}^{(n)}+\tilde{J}_{j-2}^{R}\psi_{j-2}^{(n)}+\lambda_j\psi_j^{(n)},
\label{eq:Schrodinger}
\end{equation}
where $\psi_j^{(n)}$ denotes the wave-function amplitude at site $j$. For notational simplicity, we suppress the realization index in this subsection. To characterize the localization nature of the eigenstates, we use the asymptotic Lyapunov exponent
\begin{equation}
\gamma(E_n)=\left|\lim_{N\to\infty}\frac{1}{N}\ln\left|\frac{\psi_N^{(n)}}{\psi_1^{(n)}}\right|\right|,
\label{eq:Lyapunov_def}
\end{equation}
which measures the asymptotic exponential growth or decay of the wave function.

To analytically determine the phase boundaries, we perform a non-unitary gauge transformation that maps the non-Hermitian Hamiltonian to a Hermitian counterpart. We introduce the transformed amplitudes
\begin{equation}
\psi_j^{(n)} = e^{\sum_{l=1}^{j-1} h_l}\phi_j^{(n)}.
\label{eq:GaugeTrans}
\end{equation}
Substituting Eq.~\eqref{eq:GaugeTrans} into Eq.~\eqref{eq:Schrodinger}, together with $J_j^{L}=J_1 e^{-h_j}$, $J_j^{R}=J_1 e^{h_j}$, $\tilde{J}_j^{L}=J_2 e^{-(h_j+h_{j+1})}$, and $\tilde{J}_j^{R}=J_2 e^{(h_j+h_{j+1})}$, the imaginary gauge factors cancel exactly. The spectral problem is then mapped onto the Hermitian generalized AAH model
\begin{equation}
E_n\phi_j^{(n)} = J_1(\phi_{j+1}^{(n)}+\phi_{j-1}^{(n)}) + J_2(\phi_{j+2}^{(n)}+\phi_{j-2}^{(n)}) + \lambda_j\phi_j^{(n)}.
\end{equation}

Under the gauge transformation in Eq.~\eqref{eq:GaugeTrans}, the asymptotic Lyapunov exponent defined in Eq.~\eqref{eq:Lyapunov_def} satisfies between the Lyapunov exponents of the non-Hermitian system and its Hermitian counterpart:
\begin{equation}
\gamma_{\mathrm{NH}}(E_n)
=
\left|
\lim_{N\to\infty}\frac{1}{N}\ln\left|\frac{\phi_N^{(n)}}{\phi_1^{(n)}}\right|
+
\lim_{N\to\infty}\frac{1}{N}\sum_{l=1}^{N-1}h_l
\right|
=
\left|\gamma_{\mathrm{H}}(E_n)+\overline{h}\right|,
\label{eq:Lyapunov_relation}
\end{equation}
where $\gamma_{\mathrm{NH}}$ and $\gamma_{\mathrm{H}}$ denote the Lyapunov exponents of the non-Hermitian and Hermitian systems, respectively, and $\overline{h}$ denotes the spatial average of the imaginary gauge field in the thermodynamic-limit analysis. For the Bernoulli-type disorder considered here, this average vanishes asymptotically,
\begin{equation}
\overline{h}=\lim_{N\to\infty}\frac{1}{N}\sum_{l=1}^{N}h_l=0.
\end{equation}
Therefore, the Lyapunov exponent of the non-Hermitian model is asymptotically determined by that of its Hermitian counterpart,
\begin{equation}
\gamma_{\mathrm{NH}}(E_n)\simeq \gamma_{\mathrm{H}}(E_n).
\end{equation}
For finite-size realizations, the sample-dependent average $\overline{h}_{\mathcal R}$ generally does not vanish exactly, which leads to residual non-Hermiticity under PBCs and affects the detailed spectral topology discussed in the main text. Nevertheless, for determining the localization boundary in the thermodynamic limit, $\overline{h}\to 0$, so the Lyapunov exponent is asymptotically inherited from the Hermitian auxiliary problem. Accordingly, in the present work the Lyapunov exponent should not be viewed as an ad hoc boundary-based observable for irregular ENHSE profiles, but as the asymptotic localization diagnostic inherited from the Hermitian auxiliary problem and interpreted in the original non-Hermitian frame. This mapping allows us to determine the localization transitions from known results for Hermitian AAH models. Notably, the Lyapunov exponent should not be presented as a universal statement for arbitrary non-Hermitian systems. The non-unitary imaginary-gauge transformation is available only for the physically relevant class of models considered here.

(i) For $J_2=0$, the system reduces to the standard AAH model, which exhibits a global transition at $\lambda=2J_1$. For $\lambda<2J_1$, all states are extended in the Hermitian frame, so that $\gamma_{\mathrm{H}}=0$ and hence $\gamma_{\mathrm{NH}}=0$, corresponding to the fully ENHSE phase in the non-Hermitian model. For $\lambda>2J_1$, all states are localized in the Hermitian frame, with $\gamma_{\mathrm{H}}=\ln(\lambda/2J_1)>0$, which implies $\gamma_{\mathrm{NH}}>0$ and therefore a fully localized phase.

(ii) For $J_2\neq 0$, the auxiliary Hermitian model supports an energy-dependent mobility edge $E_c$. Consequently, the Lyapunov exponent becomes energy selective: states on one side of the mobility edge satisfy $\gamma_{\mathrm{H}}(E)>0$ and hence $\gamma_{\mathrm{NH}}(E)>0$, while those on the other side satisfy $\gamma_{\mathrm{H}}(E)=0$ and hence $\gamma_{\mathrm{NH}}(E)=0$. The non-Hermitian system therefore inherits a mobility-edge boundary at the same $E_c$, but the two sectors correspond to Anderson-localized states and ENHSE-type states, respectively, rather than to localized and extended states as in the Hermitian case.

\subsection{Dynamical evolution}

We investigate the expansion dynamics of a wave packet initially localized at the lattice center, $|\Psi(0)\rangle=|j_0\rangle$, with $j_0=N/2$ for even $N$, under the non-Hermitian AAH model defined by the Hamiltonian in Eq.~\eqref{eq1}. The normalized time-evolved state is written as
\begin{equation}
|\Psi(t)\rangle = \frac{1}{\sqrt{\mathcal{N}(t)}} e^{-i\hat{H}t}|\Psi(0)\rangle,
\label{eq8}
\end{equation}
where $\mathcal{N}(t)$ is the normalization factor chosen such that $\langle \Psi(t)|\Psi(t)\rangle=1$. Accordingly, the wave-packet density satisfies $\sum_j \rho_j(t)=1$ at all times.

For numerical time propagation, we divide the total evolution time $t$ into $\bar{M}$ small steps of size $dt=t/\bar{M}$. In the limit $dt\to 0$, keeping terms up to first order in $dt$, the state at time $(\bar{m}+1)dt$ is iteratively updated as
\begin{equation}
|\Psi((\bar{m}+1)dt)\rangle=
\frac{(1-i\hat{H}dt)|\Psi(\bar{m}dt)\rangle}
{\sqrt{\langle\Psi(\bar{m}dt)|(1+i\hat{H}^{\dagger}dt)(1-i\hat{H}dt)|\Psi(\bar{m}dt)\rangle}},
\label{eq9}
\end{equation}
where $\bar{m}$ is the iteration index. For the first step, one has
\begin{equation}
|\Psi(dt)\rangle=
\frac{(1-i\hat{H}dt)|\Psi(0)\rangle}
{\sqrt{\langle\Psi(0)|(1+i\hat{H}^{\dagger}dt)(1-i\hat{H}dt)|\Psi(0)\rangle}}.
\label{eq10}
\end{equation}
After $\bar{M}$ iterations, we obtain the normalized state $|\Psi(t)\rangle$ used in the dynamical calculations.

\subsection{Disorder averaging and convergence}
Unless otherwise stated, all static disorder-averaged quantities are computed using $N_{\mathcal R}=100$ independent Bernoulli sequences of the imaginary gauge field. We verified statistical convergence by repeating representative calculations for $N_{\mathcal R}=50$, $100$, and $200$, as shown for the disorder-averaged Lyapunov exponent in Supplementary Note 4 and Supplementary Figure 4. The identified transition points remain unchanged, and the results for $N_{\mathcal R}=100$ and $200$ are nearly indistinguishable within the plotting resolution. We therefore use $N_{\mathcal R}=100$ for the static disorder-averaged quantities. For the wave-packet dynamics, where the cancellation between opposite winding sectors requires higher statistical accuracy, we use $N_{\mathcal R}=1000$ realizations for the full disorder average and $N_{\mathcal R}^{(\pm)}=1000$ realizations for each winding-resolved subset.

\section{Acknowledgements}
We thank Zhanpeng Lu and Hui Liu for fruitful discussions.

\section{Funding}
Z.X. acknowledges support from Quantum Science and Technology-National Science and Technology Major Project (Grant No. 2025ZD0300400), the National Natural Science Foundation of China under Grants No. 12375016 and No. 12461160324, and from the Beijing National Laboratory for Condensed Matter Physics under Grant No. 2023BNLCMPKF001. F.M. acknowledges support from the National Key Research and Development Program of China under Grant No. 2022YFA1404201 and the National Natural Science Foundation of China under Grant No. 12474361. G.N. acknowledges support from the National Training Program of Innovation for Undergraduates under Grant No. 202510108034.

\section{Data availability}
The data generated and analysed in this study consist of numerical eigenvalues, eigenvectors, Lyapunov exponents, winding numbers, wave-packet density profiles, and disorder-averaged observables used to produce the figures. These data are stored on institutional servers at Shanxi University. The data are available from the corresponding author Zhihao Xu upon reasonable request. Requests will be considered for academic and non-commercial research purposes, and the authors will aim to respond within four weeks.

\section{Code availability}
Custom numerical codes were used to diagonalize the tight-binding Hamiltonians, compute Lyapunov exponents, spectral winding numbers, center-of-mass observables, and wave-packet dynamics. The calculations were performed using MATLAB with standard linear-algebra routines. The parameter values used to generate each figure are provided in the figure captions and Methods. The code is stored on institutional servers at Shanxi University and is available from the corresponding author Zhihao Xu upon reasonable request. Requests will be considered for academic and non-commercial research purposes, and the authors will aim to respond within four weeks.

\section{Author contributions}
Z. X., Z. L., and F. M. conceived and designed the project. G. N. performed the numerical simulations. Z. X. provided the explanation of the numerical results. All authors contributed to the discussion of the results and wrote the paper.

\section{Competing interests}
The authors declare no competing interests.

\clearpage

\section*{Appendix}

\setcounter{equation}{0}
\setcounter{figure}{0}

\renewcommand{\theequation}{S\arabic{equation}}
\renewcommand{\thefigure}{S\arabic{figure}}

This Supplementary Information presents analyses of finite-size spectral complexification, realizations with vanishing averaged imaginary gauge field, the size robustness of the anomalous mobility edge, and the convergence with respect to disorder averaging.

\subsection{Supplementary Note 1: Finite-size-induced spectral complexification}

\begin{figure*}[htbp]
  \centering
  \includegraphics[clip, width=0.5\columnwidth]{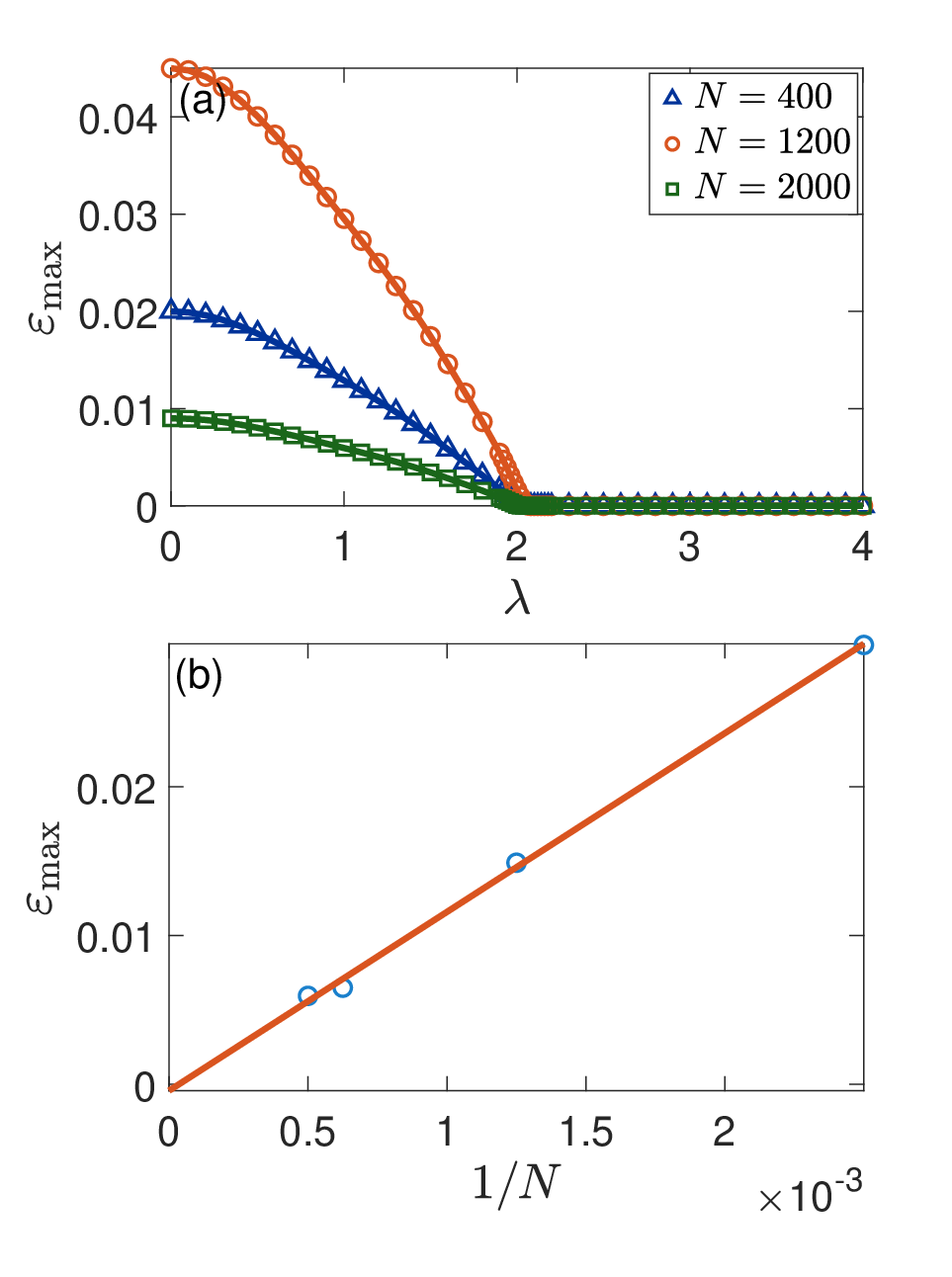}
  \caption{(a) Maximum imaginary part of the PBC eigenenergies, $\varepsilon_{\mathrm{max}}$, as a function of $\lambda$ for different system sizes $N$. (b) Finite-size scaling of $\varepsilon_{\mathrm{max}}$ at $\lambda=1$. $\varepsilon_{\mathrm{max}}$ decreases as $N$ increases and approaches zero in the thermodynamic limit. Here, $J_1=1$, $J_2=0$, and $\Delta h=0.5$.}
  \label{SFig1}
\end{figure*}

For a finite system size, the spatial average of the imaginary gauge field for a given realization $\mathcal{R}$ is defined as
\begin{equation}
\overline{h}_{\mathcal{R}}=\frac{1}{N}\sum_{j=1}^{N}h_{\mathcal{R},j}.
\end{equation}
For finite $N$, $\overline{h}_{\mathcal{R}}$ generally deviates from zero. In the ENHSE regime, this nonzero $\overline{h}_{\mathcal{R}}$ introduces a residual non-Hermiticity under PBCs and therefore gives rise to finite imaginary parts in the PBC spectrum. Consequently, the spectral complexification observed at finite $N$ originates from this finite-size effect.

To quantify this effect, we evaluate the maximum imaginary part of the PBC eigenenergies, $\varepsilon_{\mathrm{max}}$. Figure~\ref{SFig1}(a) shows $\varepsilon_{\mathrm{max}}$ as a function of $\lambda$ for $J_2=0$ and several system sizes. In the fully ENHSE phase ($\lambda<2$), $\varepsilon_{\mathrm{max}}$ remains finite for finite $N$ but decreases systematically with increasing $N$. By contrast, in the fully localized phase ($\lambda>2$), $\varepsilon_{\mathrm{max}} \approx 0$ and is nearly independent of system size. The finite-size scaling is further illustrated in Fig.~\ref{SFig1}(b) at the representative point $\lambda=1$. As $N$ increases, $\varepsilon_{\mathrm{max}}$ decreases and extrapolates to zero in the thermodynamic limit, i.e., $\varepsilon_{\mathrm{max}}\to 0$ as $N\to\infty$. This confirms that the complex PBC spectrum in the ENHSE phase is induced by finite-size effects associated with nonzero $\overline{h}_{\mathcal{R}}$. Accordingly, in the thermodynamic limit, the typical magnitude of $\overline{h}_{\mathcal{R}}$ vanishes, and the PBC spectrum becomes entirely real.

\subsection{Supplementary Note 2: Spectrum for vanishing averaged imaginary gauge field}

\begin{figure*}[htbp]
  \centering
  \includegraphics[clip, width=0.8\columnwidth]{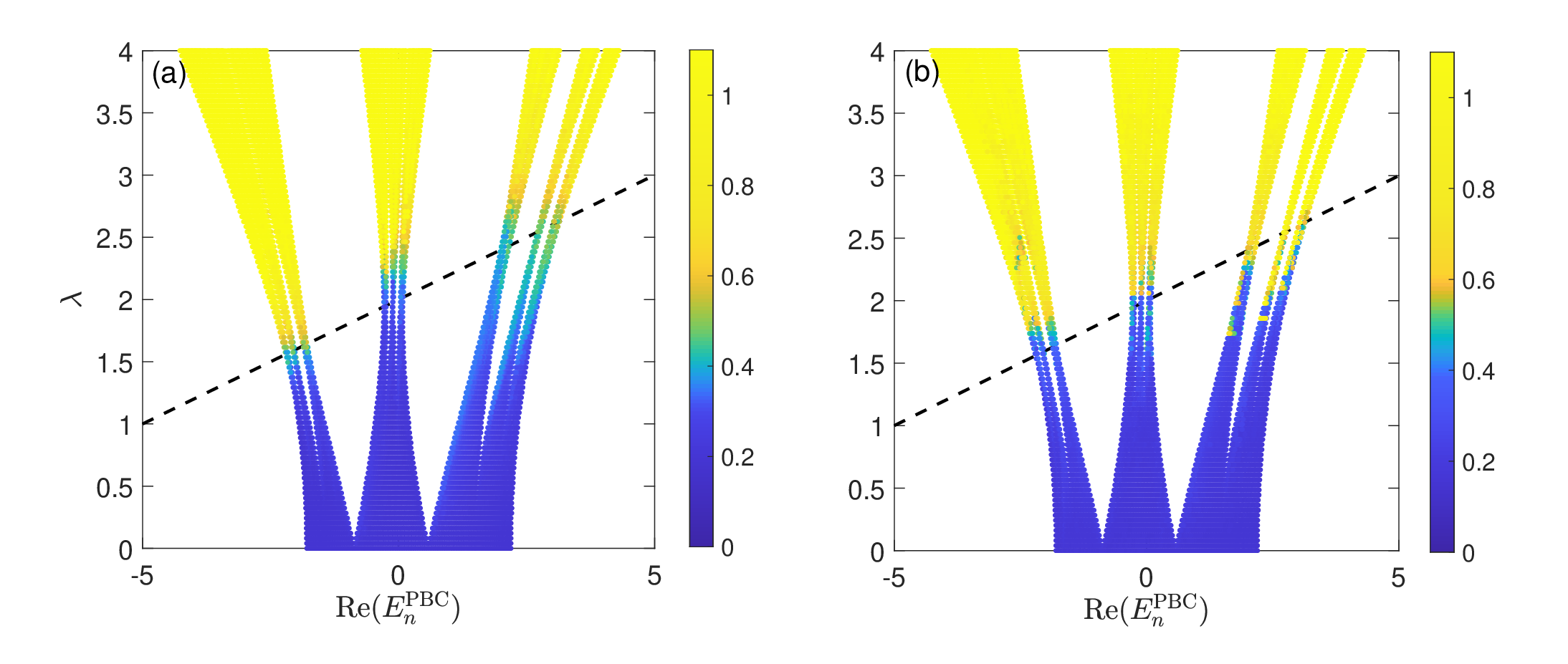}
  \caption{Spectrum for a realization with vanishing averaged imaginary gauge field, $\overline{h}_{\mathcal{R}}=0$. (a) Eigenenergy spectra under PBCs and OBCs at $\lambda=1$ for a system of size $N=300$. (b) Finite-size scaling of the maximum imaginary part of the PBC eigenenergies, $\varepsilon_{\mathrm{max}}$, at $\lambda=1$. The complex spectrum disappears when $\overline{h}_{\mathcal R}=0$, confirming that the finite imaginary parts originate from finite-size realizations with nonzero averaged imaginary gauge field. Here, $J_1=1$, $J_2=0$, and $\Delta h=0.5$.}
  \label{SFig2}
\end{figure*}

In the main text, the PBC spectrum for a finite system is shown for a representative realization $\mathcal{R}$, for which the averaged imaginary gauge field $\overline{h}_{\mathcal{R}}$ generally deviates from zero due to finite-size fluctuations. To further demonstrate that the complex eigenenergies in the ENHSE regime originate from this finite-size effect rather than being an intrinsic property of the phase, we additionally examine a realization satisfying $\overline{h}_{\mathcal{R}}=0$.

Figure~\ref{SFig2}(a) shows the eigenenergy spectra under both PBCs and OBCs at $\lambda=1$ for a realization fixing $\overline{h}_{\mathcal{R}}=0$. In this case, the PBC spectrum becomes entirely real, and the complex-energy loops present for generic finite-size realizations disappear. The finite-size behavior is further illustrated in Fig.~\ref{SFig2}(b), which shows $\varepsilon_{\mathrm{max}}$ at $\lambda=1$ under PBCs. In contrast to the case with $\overline{h}_{\mathcal{R}}\neq 0$, $\varepsilon_{\mathrm{max}}$ remains zero for all system sizes considered. These results demonstrate that the complex PBC spectrum can disappear even at finite size once $\overline{h}_{\mathcal{R}}=0$, consistent with the thermodynamic-limit behavior for generic realizations where the typical magnitude of $\overline{h}_{\mathcal{R}}$ vanishes as $N\to\infty$.

\subsection{Supplementary Note 3: Size independence of anomalous localization and mobility edges}

\begin{figure*}[htbp]
  \centering
  \includegraphics[clip, width=0.8\columnwidth]{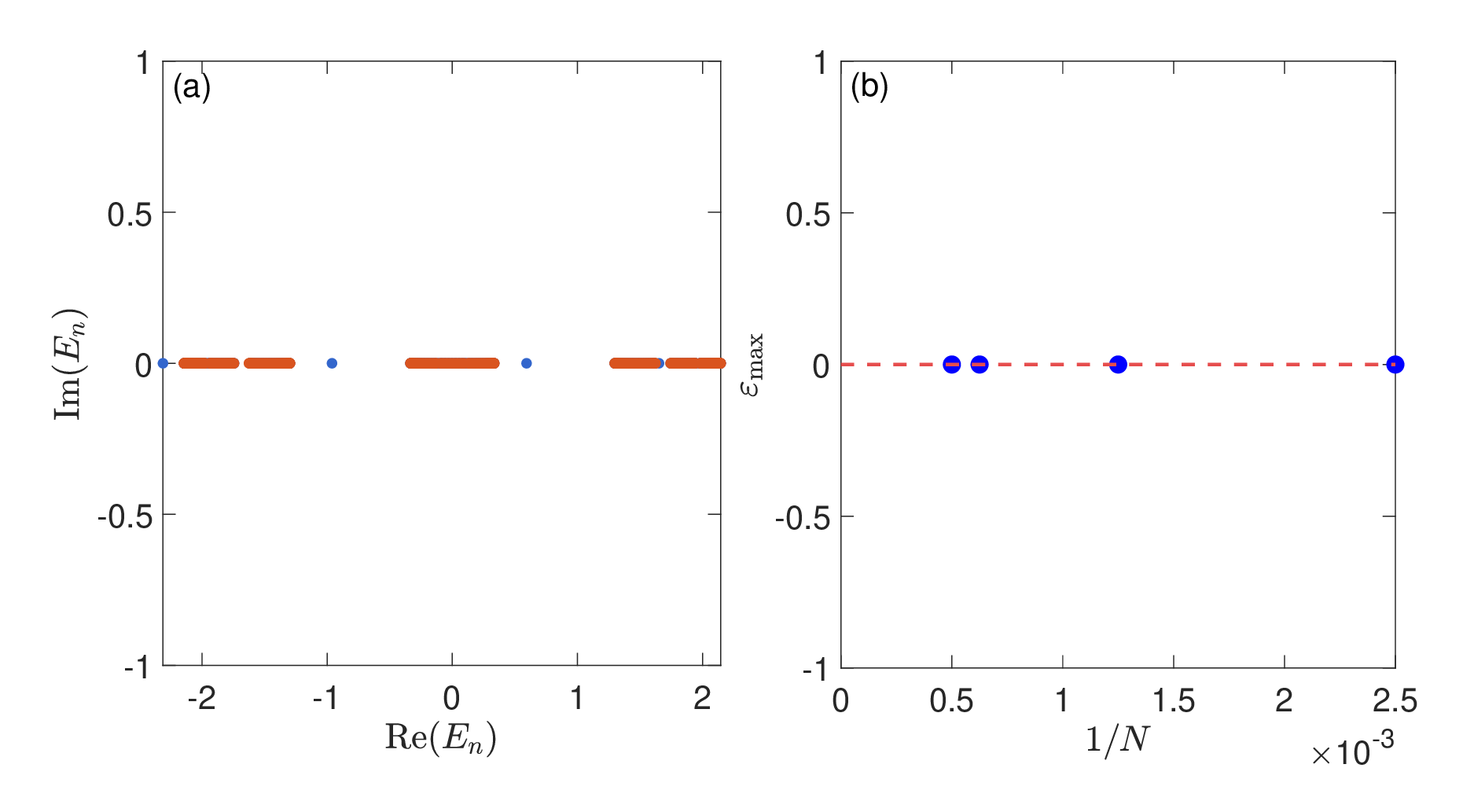}
  \caption{Localization phase diagram under PBCs in the $(\mathrm{Re}(E_n^{\mathrm{PBC}}),\lambda)$ plane, with color indicating the disorder-averaged Lyapunov exponent of the eigenstates, for (a) $N=377$ and (b) $N=610$. These results demonstrate that the anomalous localization transition is robust and independent of the specific system size. Here, $J_1=1$, $J_2=0.1$, $\Delta h=0.5$, and all data are averaged over $N_{\mathcal{R}}=100$ realizations of the imaginary gauge field.}
  \label{SFig3}
\end{figure*}

In quasiperiodic systems, the system size is often chosen to be a Fibonacci number. To examine whether our conclusions depend on this choice, we recalculate the localization phase diagram for two Fibonacci system sizes, $N=377$ and $N=610$, at fixed $J_2=0.1$ and $\Delta h=0.5$. Specifically, we compute the realization-averaged Lyapunov exponent in the $(\mathrm{Re}(E_n^{\mathrm{PBC}}),\lambda)$ plane under PBCs. As shown in Fig.~\ref{SFig3}, the resulting phase diagrams for $N=377$ and $N=610$ are both qualitatively and quantitatively consistent with that presented in the main text for $N=300$. In particular, the two localization-transition points and the position of the mobility edge remain unchanged within numerical resolution. Therefore, the characteristic features of the anomalous localization transition and the anomalous mobility edge do not depend on the specific choice of system size, and the conclusions drawn in the main text are robust for both the system sizes used there and Fibonacci-sized lattices.

\subsection{Supplementary Note 4: Convergence with respect to disorder averaging}

\begin{figure*}[htbp]
  \centering
  \includegraphics[clip, width=0.8\columnwidth]{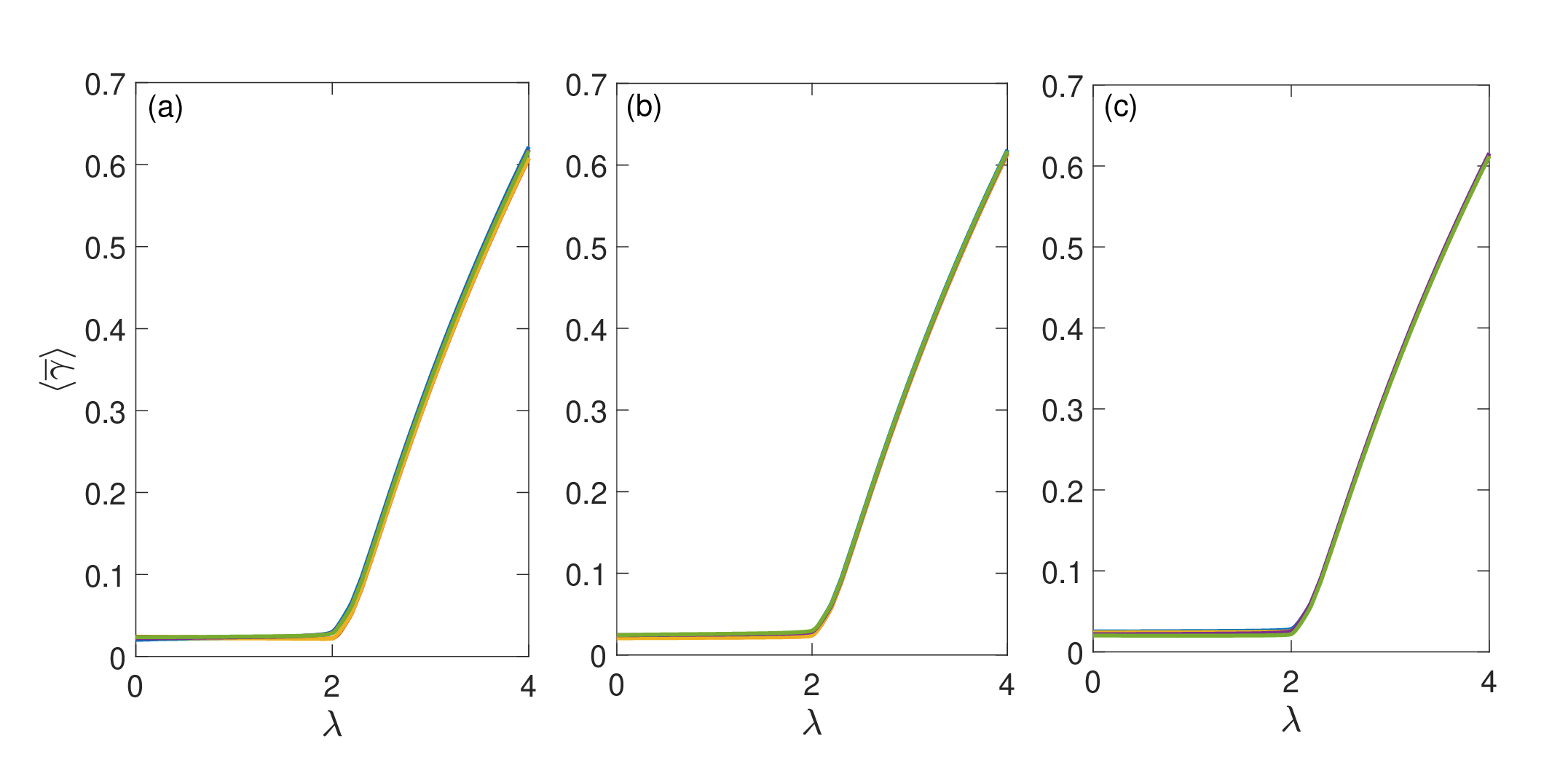}
  \caption{Realization‑averaged Lyapunov exponent $\langle \overline{\gamma} \rangle$ versus $\lambda$ for (a) $N_\mathcal{R}=50$, (b) $N_\mathcal{R}=100$, and (c) $N_\mathcal{R}=200$. Five independent trials were performed for each $N_\mathcal{R}$ to evaluate statistical fluctuations. The localization transition points coincide across all panels. Results in panels (b) and (c) are nearly indistinguishable, confirming that $N_\mathcal{R}=100$ ensures full convergence. Here, $J_1=1$, $J_2=0.1$, $\Delta h=0.5$, and $N=300$.}
  \label{SFig4}
\end{figure*}

In disordered random systems, statistical fluctuations lead to deviations in results obtained from a single disorder realization. We verify that our results and conclusions are independent of the number of disorder realizations, and demonstrate that using $N_\mathcal{R} = 100$ disorder realizations is sufficient to achieve convergence, as exemplified by calculations of the Lyapunov exponent. Figure~\ref{SFig4} presents the Lyapunov exponent obtained with different numbers of disorder realizations: (a) $N_\mathcal{R} = 50$, (b) $N_\mathcal{R} = 100$, and (c) $N_\mathcal{R} = 200$. Five independent trials were performed for each value of $N_\mathcal{R}$. The transition points are consistent across all panels, demonstrating that the results are independent of the number of disorder realizations. Nearly identical results are observed in panels (b) and (c), confirming that $N_\mathcal{R} = 100$ is sufficient to reach convergence. Since the localization properties and mobility edges in our model are directly characterized by the Lyapunov exponent, the convergence of this quantity supports the reliability of the full phase diagram presented in the main text.


\end{document}